\documentclass[twoside,twocolumn,9pt]{article}
\usepackage{extsizes}
\usepackage[super,sort&compress,comma]{natbib} 
\usepackage[version=3]{mhchem}
\usepackage[left=1.5cm, right=1.5cm, top=1.785cm, bottom=2.0cm]{geometry}
\usepackage{balance}
\usepackage{times,mathptmx}
\usepackage{sectsty}
\usepackage{graphicx} 
\usepackage{lastpage}
\usepackage[format=plain,justification=justified,singlelinecheck=false,font={stretch=1.125,small,sf},labelfont=bf,labelsep=space]{caption}
\usepackage{float}
\usepackage{fancyhdr}
\usepackage{fnpos}
\usepackage[english]{babel}
\addto{\captionsenglish}{%
  
}
\usepackage{array}
\usepackage{droidsans}
\usepackage{charter}
\usepackage[T1]{fontenc}
\usepackage[usenames,dvipsnames]{xcolor}
\usepackage{setspace}
\usepackage[compact]{titlesec}

\usepackage{nameref,zref-xr} 
\zxrsetup{toltxlabel} 
\zexternaldocument*{si_final} 
\usepackage{hyperref}

\usepackage{epstopdf}

\usepackage{siunitx}
\usepackage{isomath}
\usepackage{esint}
\usepackage{soul}
\usepackage{notes2bib}

\definecolor{cream}{RGB}{222,217,201}

\begin{document}

\pagestyle{fancy}
\thispagestyle{plain}
\fancypagestyle{plain}{

\fancyhead[C]{\includegraphics[width=18.5cm]{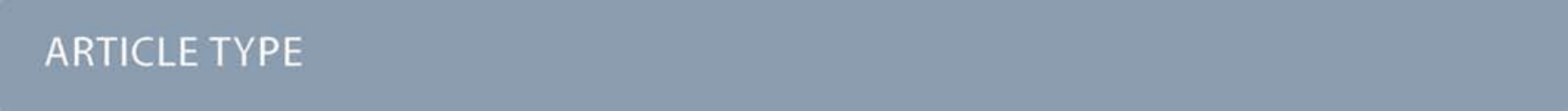}}
\fancyhead[L]{\hspace{0cm}\vspace{1.5cm}\includegraphics[height=30pt]{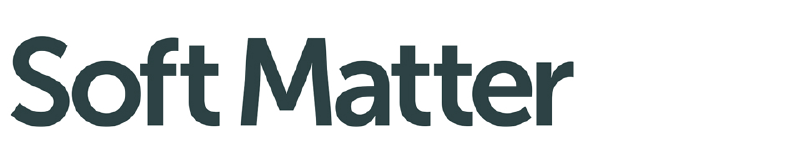}}
\fancyhead[R]{\hspace{0cm}\vspace{1.7cm}\includegraphics[height=55pt]{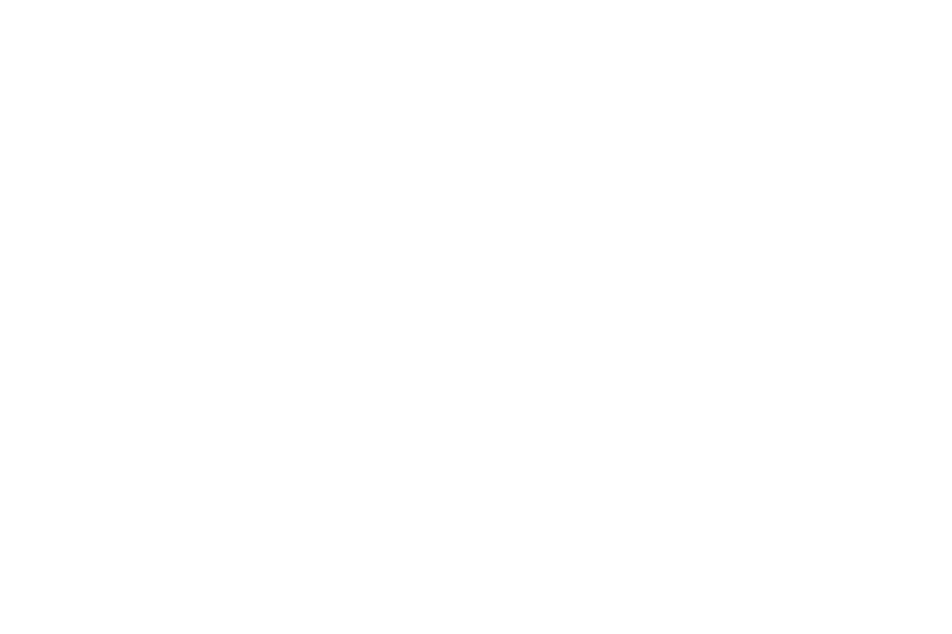}}
\renewcommand{\headrulewidth}{0pt}
}

\makeFNbottom
\makeatletter
\renewcommand\LARGE{\@setfontsize\LARGE{15pt}{17}}
\renewcommand\Large{\@setfontsize\Large{12pt}{14}}
\renewcommand\large{\@setfontsize\large{10pt}{12}}
\renewcommand\footnotesize{\@setfontsize\footnotesize{7pt}{10}}
\makeatother

\renewcommand{\thefootnote}{\fnsymbol{footnote}}
\renewcommand\footnoterule{\vspace*{1pt}%
\color{cream}\hrule width 3.5in height 0.4pt \color{black}\vspace*{5pt}} 
\setcounter{secnumdepth}{5}

\makeatletter 
\renewcommand\@biblabel[1]{#1}            
\renewcommand\@makefntext[1]%
{\noindent\makebox[0pt][r]{\@thefnmark\,}#1}
\makeatother 
\renewcommand{\figurename}{\small{Fig.}~}
\sectionfont{\sffamily\Large}
\subsectionfont{\normalsize}
\subsubsectionfont{\bf}
\setstretch{1.125} 
\setlength{\skip\footins}{0.8cm}
\setlength{\footnotesep}{0.25cm}
\setlength{\jot}{10pt}
\titlespacing*{\section}{0pt}{4pt}{4pt}
\titlespacing*{\subsection}{0pt}{15pt}{1pt}

\fancyfoot{}
\fancyfoot[LO,RE]{\vspace{-7.1pt}\includegraphics[height=9pt]{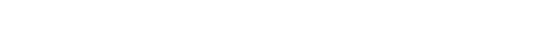}}
\fancyfoot[CO]{\vspace{-7.1pt}\hspace{13.2cm}\includegraphics{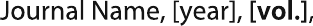}}
\fancyfoot[CE]{\vspace{-7.2pt}\hspace{-14.2cm}\includegraphics{RF}}
\fancyfoot[RO]{\footnotesize{\sffamily{1--\pageref{LastPage} ~\textbar  \hspace{2pt}\thepage}}}
\fancyfoot[LE]{\footnotesize{\sffamily{\thepage~\textbar\hspace{3.45cm} 1--\pageref{LastPage}}}}
\fancyhead{}
\renewcommand{\headrulewidth}{0pt} 
\renewcommand{\footrulewidth}{0pt}
\setlength{\arrayrulewidth}{1pt}
\setlength{\columnsep}{6.5mm}
\setlength\bibsep{1pt}

\makeatletter 
\newlength{\figrulesep} 
\setlength{\figrulesep}{0.5\textfloatsep} 

\newcommand{\topfigrule}{\vspace*{-1pt}%
\noindent{\color{cream}\rule[-\figrulesep]{\columnwidth}{1.5pt}} }

\newcommand{\botfigrule}{\vspace*{-2pt}%
\noindent{\color{cream}\rule[\figrulesep]{\columnwidth}{1.5pt}} }

\newcommand{\dblfigrule}{\vspace*{-1pt}%
\noindent{\color{cream}\rule[-\figrulesep]{\textwidth}{1.5pt}} }

\makeatother

\twocolumn[
  \begin{@twocolumnfalse}
\vspace{3cm}
\sffamily
\begin{tabular}{m{4.5cm} p{13.5cm} }

\includegraphics{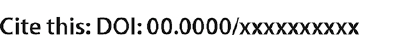} & \noindent\Large{\textbf{Crack formation and self-closing in shrinkable, granular packings$^\dag$}} \\
\vspace{0.3cm} & \vspace{0.3cm} \\

 & \noindent\normalsize{H. Jeremy Cho,\textit{$^{a}$} Nancy B. Lu,\textit{$^{a}$} Michael P. Howard,\textit{$^{b}$} Rebekah A. Adams,\textit{$^{a}$} and Sujit S. Datta$^{\ast}$\textit{$^{a}$}} \\

\includegraphics{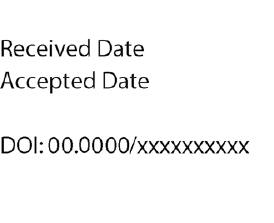} & \noindent\normalsize{Many clays, soils, biological tissues, foods, and coatings are shrinkable, granular materials: they are composed of packed, hydrated grains that shrink when dried. In many cases, these packings crack during drying, critically hindering applications. However, while cracking has been widely studied for bulk gels and packings of non-shrinkable grains, little is known about how packings of \emph{shrinkable} grains crack. Here, we elucidate how grain shrinkage alters cracking during drying. Using experiments with model shrinkable hydrogel beads, we show that differential shrinkage can dramatically alter crack evolution during drying---in some cases, even causing cracks to spontaneously ``self-close". In other cases, packings shrink without cracking or crack irreversibly. We developed both granular and continuum models to quantify the interplay between grain shrinkage, poromechanics, packing size, drying rate, capillarity, and substrate friction on cracking. Guided by the theory, we also found that cracking can be completely altered by varying the spatial profile of drying. Our work elucidates the rich physics underlying cracking in shrinkable, granular packings, and yields new strategies for controlling crack evolution.} \\

\end{tabular}

 \end{@twocolumnfalse} \vspace{0.6cm}

  ]

\renewcommand*\rmdefault{bch}\normalfont\upshape
\rmfamily
\section*{}
\vspace{-1cm}


\footnotetext{\textit{$^{a}$~Department of Chemical and Biological Engineering, Princeton University, Princeton, NJ 08544, United States. *E-mail: ssdatta@princeton.edu}}
\footnotetext{\textit{$^{b}$~McKetta Department of Chemical Engineering, University of Texas at Austin, Austin, TX 78712, United States.}}

\footnotetext{\dag~Electronic Supplementary Information (ESI) available: [details of any supplementary information available should be included here]. See DOI: 00.0000/00000000.}



\section{Introduction}
Hydrated packings of grains often crack when they dry; familiar examples of this phenomenon are cracks in mud and paint. Cracking also poses a problem in many settings: it damages structures built on clay-rich soil, causes leakage from clay barriers used for waste isolation or \ce{CO2} sequestration, alters the texture of foods, disrupts biological tissues, and limits the performance of gel particle coatings with potential for use as non-fouling films, biosensors, and drug delivery platforms \cite{goehring2015desiccation,doi:10.2113/gselements.5.2.105,Espinoza:2012iq,doi:10.1021/la050271t,doi:10.1002/admi.201500371,doi:10.1002/adfm.200902429,ARROYO20171}. In all of these cases, the individual hydrated grains \emph{shrink} when dried. However, despite its ubiquity, the influence of grain shrinkage on cracking has thus far been overlooked. Previous studies have shed light on the influence of grain size and stiffness, packing size, drying rate, capillarity, and surface adhesion \cite{doi:10.1021/la9903090,C5SM02406D,C6SM01011C,doi:10.1021/la048298k,SANTANACHCARRERAS2007160,BRINKER1990452,0034-4885-76-4-046603,PhysRevLett.91.224501,doi:10.1111/j.1151-2916.1993.tb07762.x,Giorgiutti-Dauphine2018,PhysRevLett.74.2981}, but only for bulk gels and packings of \emph{non-shrinkable} grains. Here, we demonstrate that grain shrinkage is a key factor that alters how packings crack when dried---and can be harnessed to unlock new ways to control cracking. Therefore, shrinkage should not be neglected from studies of cracking.

Experiments on model shrinkable grains reveal a range of cracking behaviors during drying: shrinkage without cracking, irreversible cracking, and most remarkably, reversible cracking in which cracks spontaneously self-close. We developed a granular model that captures these diverse cracking behaviors by connecting single-grain shrinkage and inter-grain interactions to macroscopic cracking. Moreover, we extended this granular model to a continuum model of cracking that is quantified by four universal nondimensional parameters incorporating the role of grain shrinkage, poromechanics, packing size, drying rate, capillarity, and substrate friction. This multi-scale description yields quantitative criteria to predict different cracking behaviors. Finally, guided by our theory, we found that cracking behavior can be completely altered by varying the spatial profile of drying itself, suggesting a new way to control crack evolution in shrinkable, granular packings.

\section{Results}

\subsection{Experiments reveal three different cracking behaviors}
The granular packing we consider consists of a model system of non-Brownian, cross-linked hydrogel beads held together by capillary bridges in a pendular configuration \cite{doi:10.1080/00018730600626065}. As packings in this configuration dry, bead shrinkage can cause the inter-bead capillary bridges to break, leading to the formation of cracks. Thus, our analysis starts from the moment the pendular configuration is initialized.

\begin{figure*}[h]
\centering
\includegraphics[width=\linewidth]{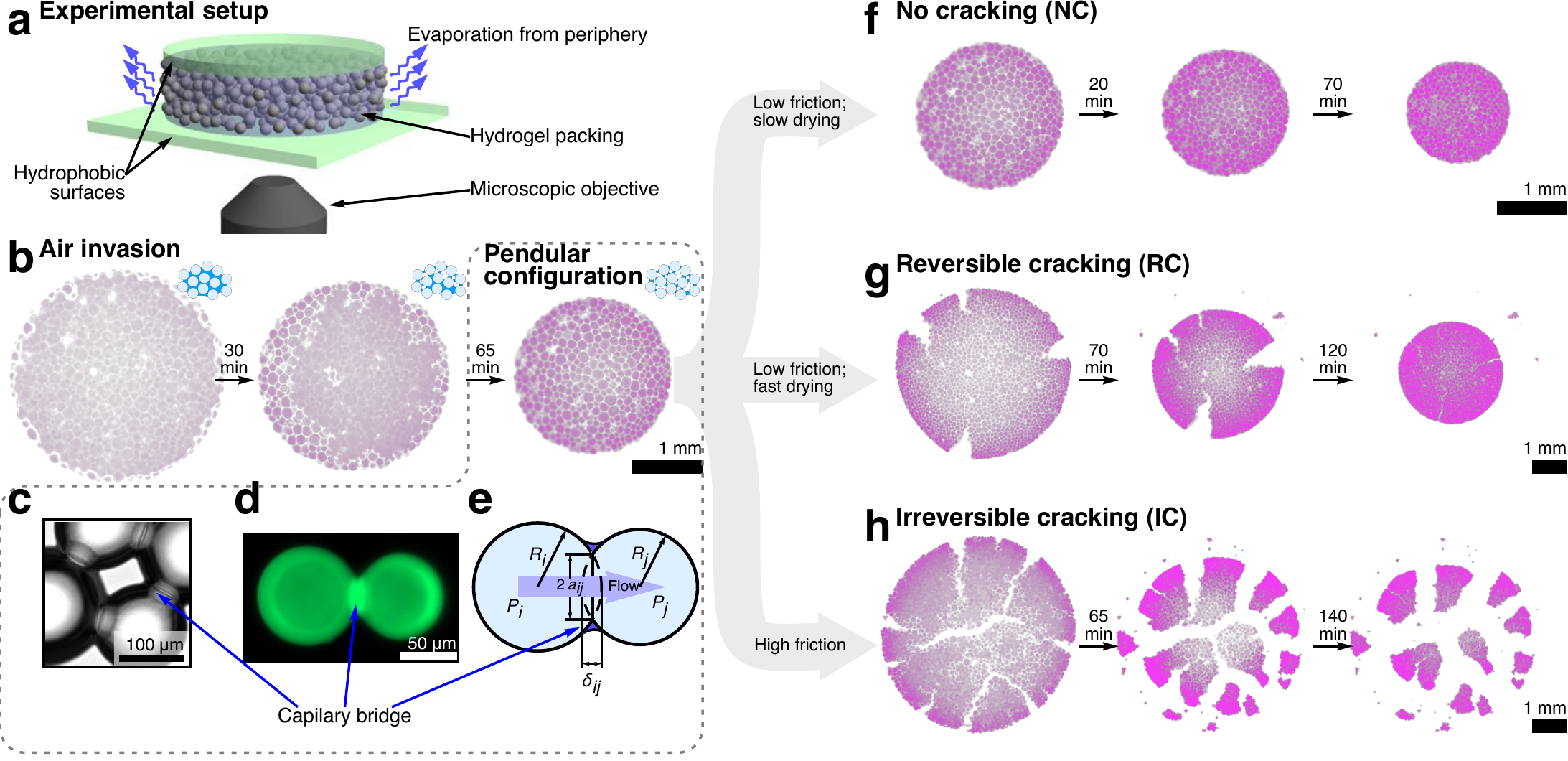}
\caption{Hydrogel disk packings crack in different ways depending on packing size. (a) A schematic of a typical drying experiment performed on an inverted microscope. (b) Time-series images and schematic diagrams show the air invasion process of a packing of hydrogel beads. Capillary forces compact the packing during this invasion process, after which, beads are in a pendular configuration with capillary bridges between them as shown in the right-most panel of (b). This pendular configuration was separately observed at higher magnification using (c) brightfield microscopy and (d) confocal fluorescence microscopy, revealing capillary bridges between beads. This capillary bridge precompresses the beads as shown by the schematic in (e). We modeled this deformation using Hertzian contact mechanics where $\delta_{ij}$ is the precompressed distance and $R$ is the radius of a bead. Flow was modeled using Darcy's law where water flows from high to low pressure, $P_i > P_j$. (f) Small packings shrink without cracking, while (g) intermediate-sized packings undergo reversible cracking where cracks self-close, and (h) large packings undergo irreversible cracking. Color shows fluorescence due to an excited dye that has diffused within the hydrogel beads; intensity increases with bead shrinkage. As shown in (g,h), cracking occurs when there is a strong radial gradient in shrinkage, while self-closing occurs when the gradient subsides as further quantified in Fig.~{\ref{fig:openclose}}.}
\label{fig:fig1}
\end{figure*}

To initialize the pendular configuration, we suspended hydrogel beads near random close packing in deionized water. Upon drying, the individual beads dehydrate and shrink in radius by a factor of 1.3 as compared to the fully-hydrated state. We confined each suspension between two smooth, hydrophobic, parallel surfaces in a disk geometry where the top surface follows the top of the packing (\hyperref[sec:matmet]{Materials and Methods}). This setup imposed azimuthal symmetry and ensured that drying to the ambient only occurred at the circular periphery as shown in Figure~\ref{fig:fig1}a; we describe more complex drying profiles in Section \ref{sec:control}. As each suspension dries, the liquid-vapor interface recedes, compacting the hydrogel beads until they are close-packed. The interstitial liquid menisci then continue to recede, leading to rapid bursts of air---known as Haines jumps---that invade the space between beads as shown and schematized in Fig.~\ref{fig:fig1}b \cite{PhysRevLett.101.094502}. In all packings tested, this air invasion results in a pendular configuration of hydrated beads interconnected by annular capillary bridges{\cite{Scheel2008}}, which we confirmed directly with brightfield and confocal fluorescence microscopy (Fig.~{\ref{fig:fig1}}c and d, respectively). These bridges precompress the individual beads by a distance $\delta$ as schematized in Fig.~{\ref{fig:fig1}}e {\cite{Moller:2007bp,Butt:2010gs,Herminghaus:2005ei}}. Our theoretical analysis starts from the moment this pendular configuration has been initialized (right-most panel of Fig.~{\ref{fig:fig1}}b); we denote the radius of the overall packing at this state as $\mathcal{R}_\text{wet}$, which is smaller than the radius of the initial suspension due to compaction during air invasion and bead precompression. Subsequent changes in the packing---such as cracking via breaking of capillary bridges---arise solely due to bead shrinkage during drying.

As packings continue to dry, we found that they evolve in one of three different ways depending on $\mathcal{R}_\text{wet}$. Packings with small $\mathcal{R}_\text{wet}$ shrink with no cracking (NC) (Fig.~\ref{fig:fig1}f, Supplementary Video~S1). For sufficiently large $\mathcal{R}_\text{wet}$, cracks form at the periphery; remarkably, for a range of $\mathcal{R}_\text{wet}$, we found that the cracks spontaneously self-close in a process we term reversible cracking (RC) to reflect the morphological similarity to the initial uncracked state (Fig.~\ref{fig:fig1}g, Supplementary Video~S2). By contrast, for even larger $\mathcal{R}_\text{wet}$, we observed irreversible cracking (IC) wherein packings break up into clusters that cannot self-close (Fig.~\ref{fig:fig1}h, Supplementary Video~S3).

\begin{figure}[h!]
\centering
\includegraphics[width=.9\linewidth]{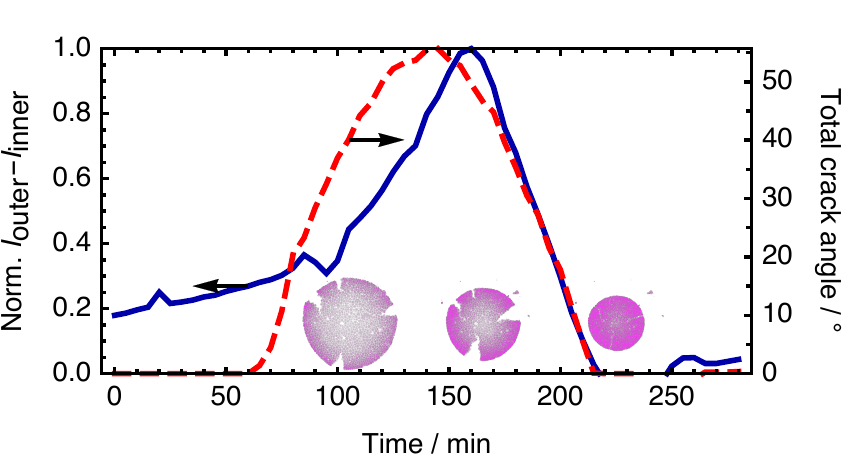}
\caption{Self-closing of cracks is concomitant with differential shrinkage subsiding. The normalized intensity difference between the inner 90\% and outer 10\% regions of the disk is plotted in blue. The sum total angles of crack (void) space at the periphery in plotted in red. Snapshots of the experiment at their respective times are also plotted. Cracks begin to close and the intensity gradient subsides after $\sim \SI{150}{\minute}. $}
\label{fig:openclose}
\end{figure}

Close inspection of the individual beads suggests the underlying mechanisms of cracking and self-closing. To quantify shrinkage during drying, we dissolved a fluorescent dye---which was subsequently absorbed by the hydrogel beads---and measured fluorescence intensity. The local dye concentration increases with bead shrinkage, leading to an increase in fluorescence intensity, as validated using direct bead size measurements (Fig.~S1). As a packing dries, we observed a differential shrinkage resulting from a mismatch in bead sizes as quantified by the gradient in fluorescence intensity in Fig.~\ref{fig:fig1}f--h and Fig.~\ref{fig:openclose}. Beads at the periphery shrink before beads in the interior, suggesting that water transport from the interior to the periphery is limited. For larger $\mathcal{R}_\text{wet}$, we observed an even larger differential shrinkage, suggesting that the rate of intra-packing water transport cannot match the rate of periphery drying; compare the radial intensity gradient between Fig.~\ref{fig:fig1}f and Fig.~\ref{fig:fig1}g. When this differential shrinkage is large enough, capillary bridges between periphery beads overstretch and break (Supplementary Video~S4), which initiates cracking. Interestingly, cracks close upon themselves when this differential shrinkage subsides (Fig.~\ref{fig:openclose}), highlighting the central importance of differential shrinkage in crack evolution. However, for very large $\mathcal{R}_\text{wet}$, while cracks still initially form at the periphery, they can also nucleate from within the packing and are unable to completely self-close, leading to immobile clusters of beads. This observation suggests that friction with the confining surfaces limits self-closing and leads to IC.

\subsection{Modeling of transport and forces}
To provide a multiphysics description of these cracking behaviors, we analyzed the physics of bead shrinkage, inter-bead water transport, and inter-bead forces during drying, as described in Sections \ref{sec:shrinkage}-\ref{sec:forces} below. We developed a discrete-element model (DEM) that fully incorporates these physics; directly solving this DEM using computer simulations yielded results that capture the experimental observations, indicating that cracking behavior can be understood using these bead-scale processes (Section \ref{sec:DEM}). Motivated by previous work on bulk gels {\cite{doi:10.1143/JPSJ.78.052001,PhysRevApplied.6.064010,Scherer:1989fe,Scherer:1992hg}}, we also extended the DEM to a simplified continuum model. In contrast to previous work, this model explicitly incorporates parameters that describe the granularity of the medium, including bead shrinkage, stiffness, permeability, and size. Importantly, however, these parameters can be collapsed into four nondimensional state variables that can predict the ultimate state of cracking (Section~\ref{sec:criteria})---providing a simple and intuitive description of cracking. Moreover, the continuum model enables us to test more complex drying profiles and how they can be harnessed to control crack evolution (Section~\ref{sec:control}).

\subsubsection{Bead shrinking}\label{sec:shrinkage}
Our experimental observations in Fig.~\ref{fig:fig1}f--h and Fig.~\ref{fig:openclose} highlight the key role of bead shrinkage in cracking. Shrinkage primarily occurs between the initial wet state---at which air invasion has completed and the packing of hydrated beads is pendular---and the final dry state (Supplementary Videos~S1--S3). We denote these states using subscripts. At the initial wet state, we assume that all hydrogel beads have the same fully hydrated volume, $V_\text{wet}$, for simplicity; conversely, at the dry state, all beads have fully shrunk to the same dry volume, $V_\text{dry}$, in equilibrium with the ambient environment. Simulations show that polydispersity does not appreciably affect our results (Supplementary Discussion and Figs.~S4--S6). We use the \emph{shrink ratio}, $S$, as a state variable to quantify the bead shrinkage between these two states:
\begin{equation}
  S \equiv \frac{R_\text{wet}}{R_\text{dry}}=\left(\frac{V_\text{wet}}{V_\text{dry}}\right)^{1/3}
\end{equation}
where $R$ and $V$ denote the radius and volume of a single bead, respectively. In nondimensional form, $\hat{R} \equiv R/R_\text{wet}$ and $\hat{V} \equiv V/V_\text{wet}$; therefore, $\hat{R}$ and $\hat{V}$ are measures of shrinkage during drying, where $\left(\hat{R}_\text{dry} = S^{-1}\right) \leq \hat{R} \leq \left(\hat{R}_\text{wet} = 1\right) $ and $\left(\hat{V}_\text{dry} = S^{-3}\right) \leq \hat{V} \leq \left(\hat{V}_\text{wet} = 1\right) $. In general, we use the hat notation ($\hat{}$) to nondimensionalize quantities by bead-scale values.

As a hydrogel bead shrinks, its hydrated polymer mesh deswells. We treat this change by assuming that the mesh size $\xi \propto V^{3/4}$, as experimentally verified by others for many hydrogels such as those used in our experiments \mbox{\cite{deGennes:1979uw,vanderSman:2015ef,ZRINYI19871139,doi:10.1021/ma011408z}}; therefore, the poromechanical properties---bulk modulus, $K$, and permeability, $\kappa$---also depend on $S$ according to power laws (as detailed in Section~{\ref{sec:scaling}}). Our DEM simulations include all of these dependencies. However, since $S=1.3$ is of order unity in the experiments, these dependencies are weak; an example demonstrating this is given in Fig.~{\ref{fig:modelvalfig}}. Hence, for simplicity, we focus on a version of the continuum model that does not consider size-dependent changes in bead properties in the following discussion.

\subsubsection{Water transport and evaporation}\label{sec:transport}
The experiments shown in Fig.~\ref{fig:fig1}f--h and Fig.~\ref{fig:openclose} highlight the importance of intra-packing water transport and periphery drying on crack behavior. In the experiments, beads inside the packing do not shrink immediately after air invasion (Fig.~{\ref{fig:fig1}}b)---rather, the overall packing slightly compacts due to bead precompression by capillary bridges---indicating that the partial pressure of water vapor is near saturation in the air-invaded region. We therefore assume that water transport through the vapor region is negligible. Water transport at the single-bead scale can then be described using Darcy's law: $\vectorsym{q} = -\left(\kappa/\mu\right)\nabla P$ where $\vectorsym{q}$ is the liquid flux, $\mu$ is the dynamic viscosity of water, and $P$ is the liquid pore pressure. Periphery drying imposes a gradient of $P$ within the packing, leading to inter-bead water transport. This evaporation-induced flow is similar to that found in drying suspensions and emulsions \mbox{\cite{Deegan1997,C2SM27285G}}; however, in our packings, flow is limited by the permeability of the hydrogel beads themselves \mbox{\cite{Scherer:1992hg}}. This behavior is quantified by the poroelastic diffusion coefficient, $D \equiv \frac{\kappa K}{\mu \hat{V}}$ (Section~\mbox{\ref{sec:diffusion}}), which converts the pressure gradient to a gradient in bead volumes \mbox{\cite{goehring2015desiccation,PhysRevApplied.6.064010}}. Thus, our DEM simulations use a discretized form of Darcy's law that explicitly considers differences in bead size---that is, differential shrinkage---as detailed in Section~{\ref{sec:demflow}}.

To develop a continuum version of this granular model, we assume that there is a continuum of beads within a packing. We thus treat the system as having a shrinkage field of $\hat{V}\left(\tilde{r}\right)$, which specifies $\hat{V}_i$ for a grain $i$ located at $\tilde{r}$---the Lagrangian radial coordinate normalized by time-dependent packing radius ($\tilde{r} = 1$ at the packing periphery). We use the tilde notation ($\tilde{}$) to nondimensionalize quantities by macroscopic values. Applying fluid mass conservation yields a diffusion equation governing the time-dependent distribution of water, as derived in Section~\ref{sec:diffusion}:
\begin{equation}
\frac{\partial \hat{V}}{\partial t} = D_\text{wet} \nabla^2 \hat{V} \text{.}
\label{eqn:simpletransport}
\end{equation} 
where $t$ is time. Here, the poroelastic diffusion coefficient is defined similarly to the granular case, but is constant and evaluated at the wet state: $D_\text{wet} = \kappa_\text{wet} K_\text{wet}/\mu $. The simplifying assumptions described above do not considerably affect the model results, as exemplified in Fig.~{\ref{fig:modelvalfig}}. However, they enable us to analytically solve for the differential shrinkage in the packing that arises during drying.

We model periphery drying using a convective boundary condition with a mass transfer coefficient $h$, such that the volumetric flux of water due to evaporation $j_\text{evap}$ is proportional to the concentration of water at the periphery: $j_\text{evap} = h \left(\hat{V}-\hat{V}_\text{dry}\right)$. In nondimensional form, this condition is expressed as
\begin{equation}
  -\left. \left(\frac{\partial \psi}{\partial \tilde{r}}\right)\right|_{\tilde{r}=1} = \underbrace{\frac{h \mathcal{R}_\text{wet}}{D_\text{wet}}}_\text{Bi}\psi 
\label{eqn:bc}
\end{equation}
where $\psi \equiv \left(\hat{V}-\hat{V}_\text{dry}\right)/\left(\hat{V}_\text{wet}-\hat{V}_\text{dry}\right)$ is the fractional water content. The nondimensional quantity $\text{Bi}$ is the Biot number, which quantifies the rate of periphery drying ($\sim h/\mathcal{R}_\text{wet}$) relative to the rate of intra-packing water transport ($\sim D_\text{wet}/\mathcal{R}_\text{wet}^2$). We therefore expect that the packing shrinks more non-uniformly when $\text{Bi} \gg 1$ and more uniformly when $\text{Bi} \ll 1$. We confirmed this expectation by analytically solving our simplified continuum diffusion equation (Eq.~{\ref{eqn:simpletransport}}) with this convective boundary condition (Eq.~{\ref{eqn:bc}}), yielding a full solution for $\psi$ that exhibits greater differential shrinkage at higher $\text{Bi}$:
\begin{equation}
  \psi = \sum_{n=1}^\infty C_n e^{-\lambda_n^2 \tau} J_0\left(\lambda_n \tilde{r}\right)
  \label{eqn:simpletransportsolution}
\end{equation}
where $C_n \equiv \frac{2 J_1\left(\lambda_n\right)}{\lambda_n\left(J_0^2\left(\lambda_n\right)+J_1^2\left(\lambda_n\right)\right)}$, $J_0$ and $J_1$ are the zeroth and first-order Bessel functions of the first kind, $\tau \equiv D_\text{wet} t/\mathcal{R}_\text{wet}^2$ is nondimensional time expressed as the Fourier number, and $\lambda_n$ are the roots of $\lambda_n\frac{J_1\left(\lambda_n\right)}{J_0\left(\lambda_n\right)} = \text{Bi}$. The $\lambda_n$ dependence on $\text{Bi}$ confirms the expectation that higher $\text{Bi}$ results in a greater differential shrinkage---as quantified by a difference in $\psi$ between the periphery and interior---throughout the drying process; this relationship is plotted in Fig.~S2. Furthermore, using both our DEM simulations and continuum approximation, we will show that the stresses that develop due to differential shrinkage at high $\text{Bi}$ lead to an increased propensity for cracking. This expectation is consistent with our experiments in Fig.~\ref{fig:fig1} where increased $\mathcal{R}_\text{wet}$---which increased $\text{Bi}$---led to more cracking. $\text{Bi}$, along with $S$, is therefore a state variable that governs cracking behavior.

\subsubsection{Force interactions}\label{sec:forces}
To model the stresses that develop due to differential shrinkage during drying, we start by describing the contact and capillary forces between beads in the pendular configuration. We neglected inter-bead friction since this has been measured to be negligible \cite{doi:10.1063/1.4983047}. We also first consider the case when friction with the substrate is negligible; however, we explicitly incorporated these interactions in the DEM simulations, as described below. As shown in Fig.~\ref{fig:fig1}c,d and schematized in Fig.~\ref{fig:fig1}e, the balance of contact and capillary forces precompresses beads by an amount $\delta_{ij}$. Motivated by previous work \mbox{\cite{Moller:2007bp,Fogden:1990ej,Butt:2010gs,Herminghaus:2005ei}}, and validated for hydrogel beads \mbox{\cite{PhysRevE.84.011302,PhysRevE.82.041403}}, we modeled the contact force using a simple Hertzian mechanical description: $F_{\text{con},ij} =\left(4/3\right)\bar{K}_{ij} \bar{R}_{ij}^{1/2} d_{ij}^{3/2}$ where $\bar{K}_{ij}$ is the effective bead bulk modulus, $\bar{R}_{ij}$ is the effective bead size, and $d_{ij}$ is the instantaneous deformation distance. Motivated by our observations of thin annular capillary bridges (Fig.~{\ref{fig:fig1}}c,d) and previous work \mbox{\cite{Moller:2007bp,doi:10.1021/la000657y}}, we modeled the capillary force as $F_{\text{cap},ij} = 2\pi a_{ij} \gamma$, where $\gamma$ is the surface tension and the capillary bridge radius is equivalent to the contact radius $a_{ij}$. At mechanical equilibrium when the capillary force is equivalent to the contact force, $d_{ij}=\delta_{ij} = 3\pi \gamma / \left(2\bar{K}_{ij} \right)$. Our DEM simulations explicitly incorporate the variation of this equilibrium compression distance with bead shrinkage (Section~{\ref{sec:scaling}}), and its value at the wet state provides a way to characterize the combined influence of capillary and elastic forces. In nondimensional form, this parameter is given by (Section~\ref{sec:demforces}):
\begin{equation}
  \hat{\delta}_\text{wet} \equiv \frac{\delta_\text{wet}}{R_\text{wet}} = \frac{8 \pi \gamma}{3 K_\text{wet} R_\text{wet}} = \frac{4}{3\hat{K}_\text{wet}}
  \label{eqn:deltawet}
\end{equation}
where we nondimensionalize the bulk modulus $K$ by a characteristic capillary pressure $2\pi\gamma/R_\text{wet}$. We expect that packings composed of softer beads---with high $\hat{\delta}_\text{wet}$---are more crack-resistant since they require a larger separation between bead centers to break capillary bridges. Our DEM simulations confirm this expectation as we will show later. $\hat{\delta}_\text{wet}$, along with $S$ and $\text{Bi}$, is therefore a state variable that governs cracking behavior.

\begin{figure}[h!]
\centering
\includegraphics[width=\linewidth]{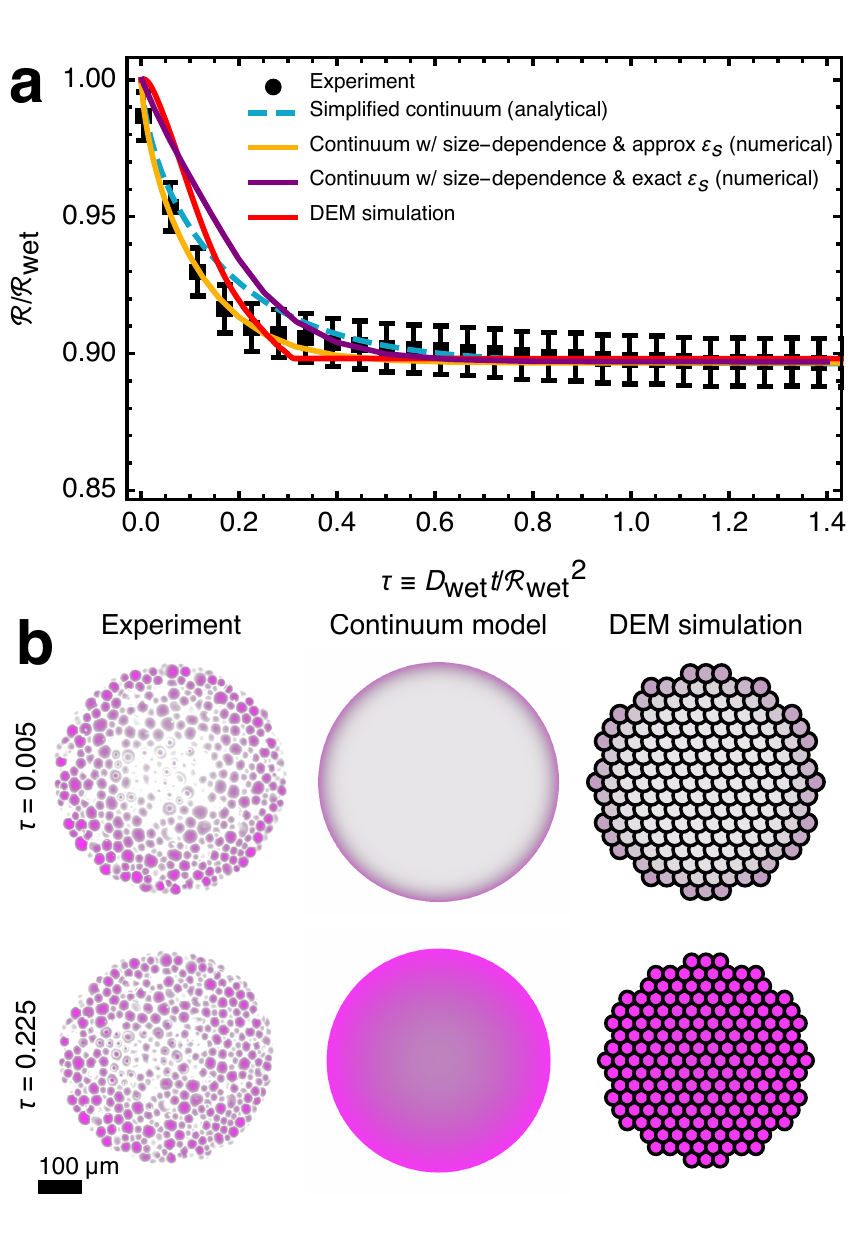}
\caption{An experiment of a drying packing with NC behavior supports our continuum model and DEM simulations. (a) Time-dependent packing size for experiments and models where time is quantified by the nondimensional Fourier number ($\tau \equiv D_\text{wet} t / \mathcal{R}_\text{wet}^2$). The blue, dashed curve represents the simple analytical model (Eqs.~{\ref{eqn:simpletransport}},{\ref{eqn:analyticalur}}); this model assumes constant bead properties (set to the wet state values), $\mathcal{V}/V_\text{wet} \approx 1$ (Section~{\ref{sec:transport}}), and a linear dependence of $\epsilon_\text{s}$ on $\hat{V}$ (Eq.~{\ref{eqn:shrinkagestrain}}). The gold curve takes into account bead size dependent property changes via polymer scaling laws (numerical solutions to Eq.~{\ref{eqn:fullcontinuumappendix}} with linear elasticity) and incorporates a time-dependent $\mathcal{V}/V_\text{wet}$. The purple curve additionally incorporates the nonlinear dependence on $\hat{V}$ (Eq.~{\ref{eqn:actualshrinkagestrain}}). Neither of the complexities incorporated to obtain the gold and purple curves change the results greatly. The red curve is a DEM simulation. (b) Plots of shrinkage in the constant diffusion coefficient continuum model (middle) and DEM simulation (right) at two time points in comparison to experiment (more purple indicates smaller bead size). Input parameters into the models were set to experimentally determined values ($\hat{\delta}_\text{wet} = 0.35$, $S = 1.3$). There are no fitting parameters in the models.}
\label{fig:modelvalfig}
\end{figure}

To determine the stresses due to differential shrinkage, we extend this granular description to a continuum model using linear elasticity. A linearization of contact and capillary forces in granular packings yields an effective Young's modulus \cite{Moller:2007bp},
\begin{equation}
  \mathcal{E} = \left(3/16\right)\sqrt{3 K_\text{wet} \gamma / \left(\pi R_\text{wet}\right)} \text{.}
  \label{eqn:effmodulus}
\end{equation}
Using this modulus, we build a constitutive relationship between stress, $\tensorsym{\sigma}$, and strain, $\tensorsym{\epsilon}$, by adding an extra drying-induced shrinkage term to Hooke's law:
\begin{equation}
  \tensorsym{\epsilon} = \underbrace{\frac{1+\nu}{\mathcal{E}}\tensorsym{\sigma} - \frac{\nu}{\mathcal{E}}\text{tr}\left(\tensorsym{\sigma}\right)\matrixsym{I}}_\text{strain from stress} + \underbrace{\epsilon_\text{s}\matrixsym{I}}_\text{strain from shrinkage}
  \label{eqn:elasticity1}
\end{equation}
where the quantity $\epsilon_\text{s}<0$ represents the strain due to bead shrinkage from water evaporation. At the wet state, $\epsilon_\text{s} = 0$ and at the dry state, $\epsilon_\text{s} = \epsilon_\text{dry}$, which can be explicitly related to the state variables $S$ and $\hat{\delta}_\text{wet}$ (Eq.~\ref{eqn:edry}). We found that this quantity can be approximated by
\begin{equation}
  \epsilon_\text{s} \approx \epsilon_\text{dry} \left(1 - \psi\right) \text{.}
  \label{eqn:shrinkagestrain}
\end{equation}
as detailed in Section~\ref{sec:continuumshrinkage}. This simplifying assumption does not considerably affect the model results, as exemplified in Fig.~{\ref{fig:modelvalfig}}. Moreover, incorporating this approximation into the constitutive relation (Eq.~{\ref{eqn:elasticity1}}), along with the full solution for water transport throughout the packing (Eq.~\ref{eqn:simpletransportsolution}), yields the full, analytical time-dependent displacement and azimuthal stress fields $\tilde{u}_r$ and $\sigma_{\theta\theta}$, respectively:
\begin{equation}
  \tilde{u}_r = \epsilon_\text{dry}\left(\tilde{r}-\sum_n^\infty \frac{2 C_n}{3 \lambda_n}\left(\tilde{r} J_1\left(\lambda_n\right)+ 2 J_1\left(\lambda_n \tilde{r}\right)\right) e^{-\lambda_n^2 \tau} \right) \label{eqn:analyticalur}\text{,}
\end{equation}
\begin{equation}
  \frac{\sigma_{\theta\theta}\left(\tilde{r},\tau\right)}{\mathcal{E}} = -\frac{\epsilon_\text{dry}}{\tilde{r}}\sum_n^\infty \frac{C_n}{\lambda_n} \left(\tilde{r} J_1\left(\lambda_n\right)+  J_1\left(\lambda_n \tilde{r} \right) - \tilde{r} \lambda_n J_0\left(\lambda_n \tilde{r}\right) \right) e^{-\lambda_n^2 \tau} \text{.}
  \label{eqn:stress}
\end{equation}
This time-dependent solution depends directly on the state variables: $S$ and $\hat{\delta}_\text{wet}$ through $\epsilon_\text{dry}$ (Eq.~\ref{eqn:edry}), and $\text{Bi}$ through $C_n$ and $\lambda_n$ (Eq.~\ref{eqn:simpletransportsolution}). Specifically, the shrinkage---as quantified by $u_r$---and the stress increase with $\text{Bi}$ as shown in Fig.~S2. We tested this continuum model by direct comparison with a NC drying experiment, which most closely represents the radial geometry of our continuum model at the same $S$, $\text{Bi}$, and $\hat{\delta}_\text{wet}$; we found excellent agreement in shrinkage between the two as shown in Fig.~\ref{fig:modelvalfig}.

The continuum solution of stress (Eq.~\ref{eqn:stress}) reveals how the state variables $S$, $\text{Bi}$, and $\hat{\delta}_\text{wet}$, which naturally arise from the granular picture, govern the onset of macroscopic cracking. As drying progresses, differential shrinkage grows, causing the stress to develop and eventually reach a maximum near the onset of cracking. After this maximum, the stress diminishes as the interior of the packing begins to shrink, reducing the amount of differential shrinkage. This peaking behavior is also captured mathematically in Eq.~{\ref{eqn:stress}}---the stress is a series of decaying time exponentials that could be positive or negative. Although this maximum, $\sigma_\text{max}$, cannot be expressed in closed form, we found that it can be closely approximated by
\begin{equation}
  \tilde{\sigma}_\text{max} \equiv \frac{\sigma_\text{max}}{\mathcal{E}} = -\epsilon_\text{dry}\left(1+\frac{10}{\text{Bi}} \right)^{-1}
  \label{eqn:genericstress}
\end{equation}
which increases with $\text{Bi}$ sigmoidally (Fig. S3), as well as with increasing $S$ and decreasing $\hat{\delta}_\text{wet}$ via $\epsilon_\text{dry}$ (Eq.~\ref{eqn:edry}). Therefore, our continuum model can quantify how a packing is more likely to crack when it is rapidly dried (high $\text{Bi}$) and composed of more shrinkable (high $S$) and stiffer beads (low $\hat{\delta}_\text{wet}$).

\begin{figure*}[h!]
\centering
\includegraphics[width=\linewidth]{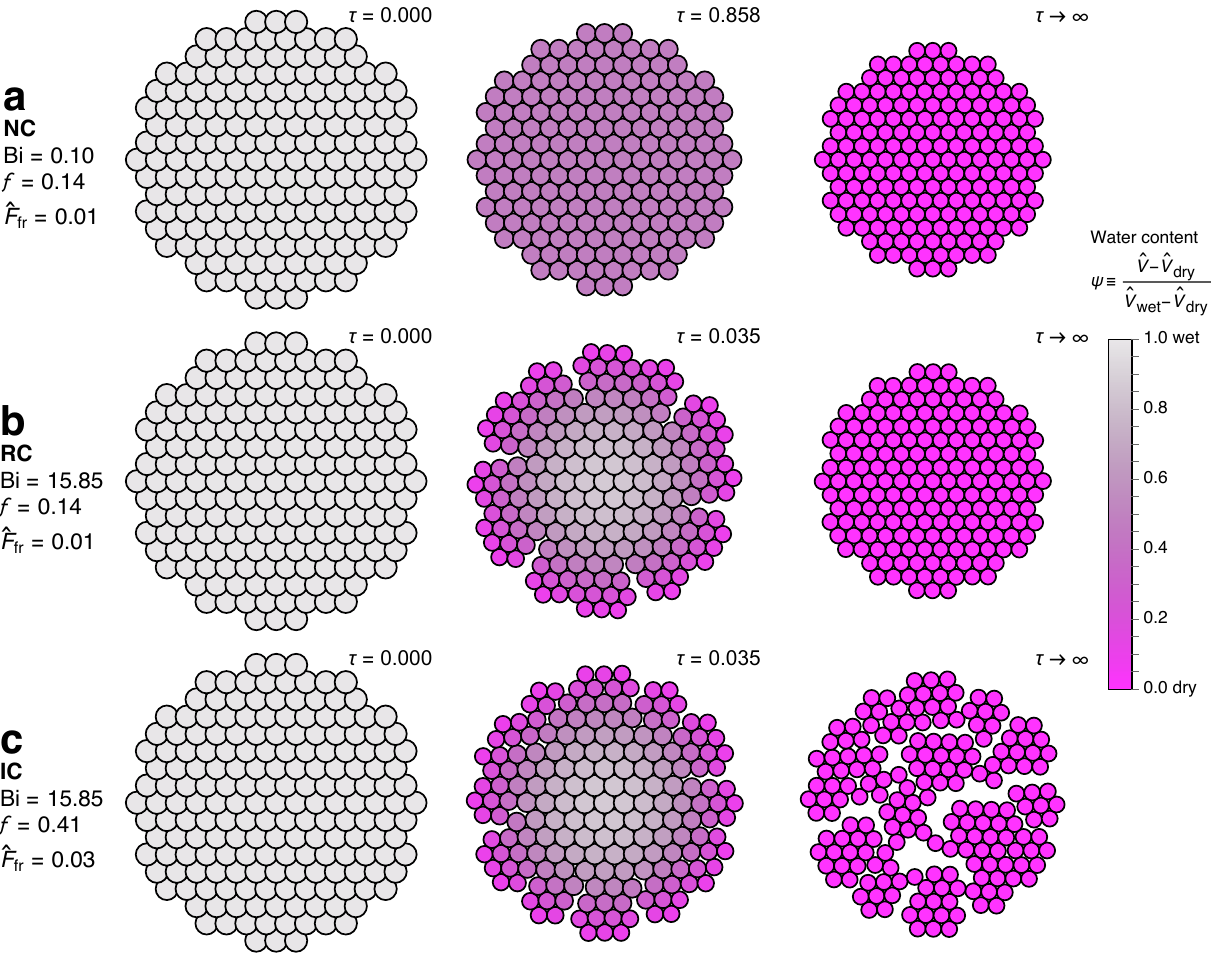}
\caption{DEM simulations can reproduce all three cracking behaviors. (a) At low $\text{Bi}$ and low $f$ a simulated packing shrinks uniformly---as indicated by the absence of color gradient---and no cracking (NC) results as bonds do not break over time, which is quantified using the Fourier number ($\tau \equiv D_\text{wet} t / \mathcal{R}_\text{wet}^2$). (b) At a higher $\text{Bi}$ but the same $f$ compared to (a), bonds break near the periphery forming cracks. Eventually, the beads arrange themselves to self-close these cracks in a process of reversible cracking (RC). (c) By increasing $f$, cracks develop similarly to (b); however, due to the decreased mobility of beads from higher friction, the packing breaks up, resulting in irreversible cracking (IC). In all cases, $S = 1.4$, $\hat{\delta}_\text{wet} = 0.22$, and $M = 187$. This data set is also provided as Supplementary Videos~S5--S7.}
\label{fig:demresultsqual}
\end{figure*}

\subsection{DEM simulations capture all three cracking behaviors}\label{sec:DEM}
We implemented the granular physics described heretofore in DEM simulations. Our DEM simulation method solves for bead sizes, bead positions, and capillary-bridge bonds using the granular forces and water transport rules we derived. The method incorporates a gradient-descent scheme to find the mechanical minimum for a packing of beads with time-dependent sizes. The packing is initially hexagonal-close-packed, but we found that this condition does not appreciably affect our results (Supplementary Discussion, Figs. S4--S6). A hybrid event-detection scheme breaks bonds when they become overstretched. We tested this time-dependent DEM simulation by direct comparison with a NC drying experiment at the same $S$, $\text{Bi}$, and $\hat{\delta}_\text{wet}$; we found excellent agreement in shrinkage between the two, and with the continuum model, as shown in Fig.~\ref{fig:modelvalfig}. To incorporate friction with the substrate into the DEM simulations, we first determine the net capillary and contact forces on each individual bead. If this net force exceeds a constant, static friction threshold, $F_\text{fr}$, then the bead moves; otherwise, it does not.

\begin{figure*}[h!]
\centering
\includegraphics[width=\linewidth]{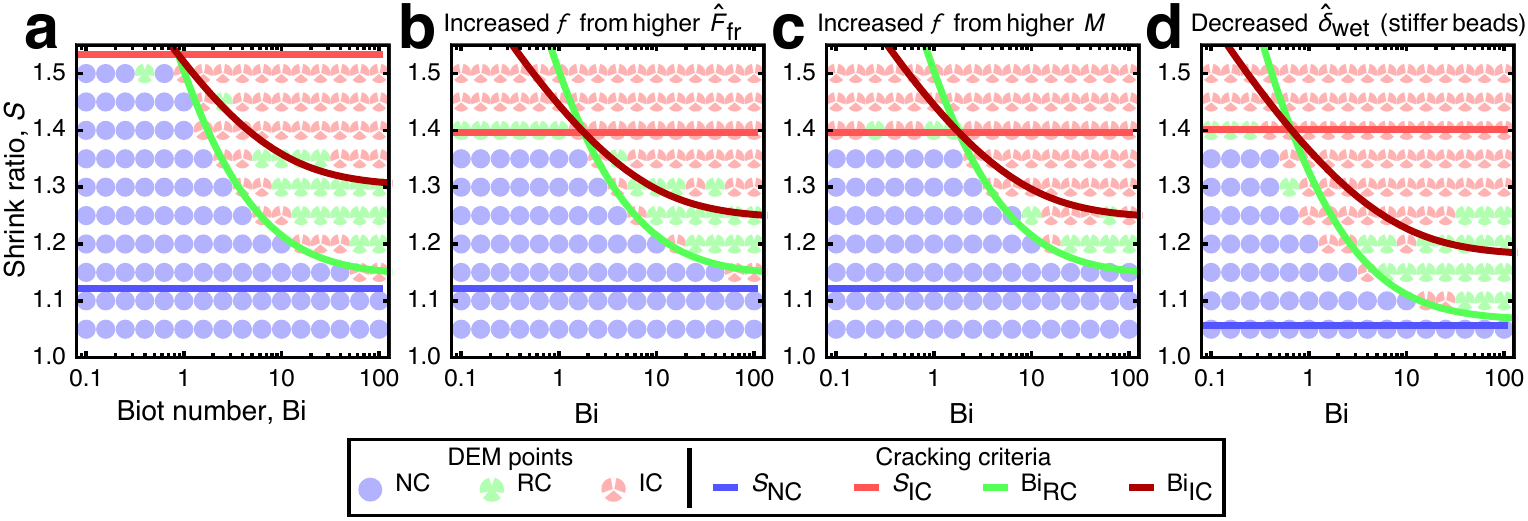}
\caption{State diagrams of DEM simulation results show all three cracking behaviors: no cracking (NC), reversible cracking (RC), and irreversible cracking (IC). The state variables are $S$, $\text{Bi}$, $\hat{\delta}_\text{wet}$, and $f$. Cracking behaviors were classified according to the number of bond breaks and the distances between beads at the dry state (\ref{sec:demcracking}). The occurrence of some data points being inconsistent with cracking criteria can be attributed to an artifact of arbitrary classification. Compared to (a) the low $f$ case, increasing $f$ by (b) the static friction threshold, $F_\text{fr}$, or (c) the packing size, $M$, results in nearly identical increases in the propensity for IC. (d) Stiffening the beads---lower $\hat{\delta}_\text{wet}$---results in a higher propensity for both RC and IC. Solid curves show boundaries between different cracking behaviors determined using our continuum model (Eqs.~\ref{eqn:BiRC},\ref{eqn:SIC},\ref{eqn:BiIC},\ref{eqn:Smin}).}\label{fig:fig4}
\end{figure*}

To test the influence of $S$, $\text{Bi}$, and $\hat{\delta}_\text{wet}$ on cracking as suggested by our continuum model, we ran several thousand simulations varying these state variables. We were able to classify all results as either NC, RC, or IC (Fig.~\ref{fig:demresultsqual}, Supplementary Videos~S5--S7) as indicated by blue, green, and red symbols in the state diagrams in Fig.~\ref{fig:fig4}. Thus, the experimentally-observed cracking behaviors can be captured using the bead-scale interactions that are incorporated in the DEM. Moreover, clear boundaries delineate these different behaviors. Cracking results (RC and IC) have higher $S$, higher $\text{Bi}$, and lower $\delta_\text{wet}$ compared to NC results. Thus, the boundary describing the onset of cracking agrees with the expectations of our continuum model.

Our experiments show that for large packings, clusters of beads can be immobilized and cause IC; this observation suggests that friction with the substrate also governs cracking behavior. To incorporate the observed dependence on packing size, we independently varied the number of beads that make up the area of the packing, $M$. We found that increasing either $F_\text{fr}$ or $M$ increases the propensity for IC as shown in Fig.~\ref{fig:fig4}. In particular, an increase in $F_\text{fr}$ or $M$ moves the NC--IC and RC--IC borders---denoted by $S_\text{IC}$ and $\text{Bi}_\text{IC}$, respectively---to lower $S$ and lower $\text{Bi}$. Both $F_\text{fr}$ and $M$ move these borders in similar ways, suggesting that there is a characteristic friction for the entire packing that depends on both of these parameters.

This characteristic friction can be determined by examining the force required to break up the packing into smaller clusters $N$ beads across. Breakup occurs when the friction force that immobilizes a cluster, $\sim N \hat{F}_\text{fr}$, balances the capillary force holding the boundary of the cluster to the packing, $\sim N^{1/2} H \hat{F}_\text{cap}$; $H$ is the number of monolayers quantifying the packing height, $\sim N^{1/2} H$ is the number of beads at the cluster boundary, and the single-bead forces $\hat{F}_\text{fr}$ and $\hat{F}_\text{cap}$---assumed to be uniform throughout the packing for this simple estimate---are nondimensionalized by $2\pi\gamma R_\text{wet}$. At the boundary of IC, $N \sim M$; thus, to incorporate both packing size dependence and single-bead friction within a characteristic friction for the entire packing, we can define
\begin{equation}
  f \equiv \frac{M^{1/2} \hat{F}_\text{fr}}{H}\text{.}
  \label{eqn:f}
\end{equation}
The $f$ parameter can be interpreted as the force per grain to immobilize clusters of $M$ beads in the limit of low $\text{Bi}$ since we assumed $\hat{F}_\text{cap}$ to be uniform.  DEM simulations with different $M$ and $F_\text{fr}$, but the same value of $f$, yield nearly identical results as shown in Fig.~\ref{fig:fig4}b,c and Fig. S7. This finding confirms our expectation that a single parameter $f$ collapses the effects of varying $M$ and $\hat{F}_\text{fr}$. As such, $f$ is also a state variable that governs cracking, along with $S$, $\text{Bi}$, and $\hat{\delta}_\text{wet}$.

\begin{figure*}[h!]
\centering
\includegraphics[width=\linewidth]{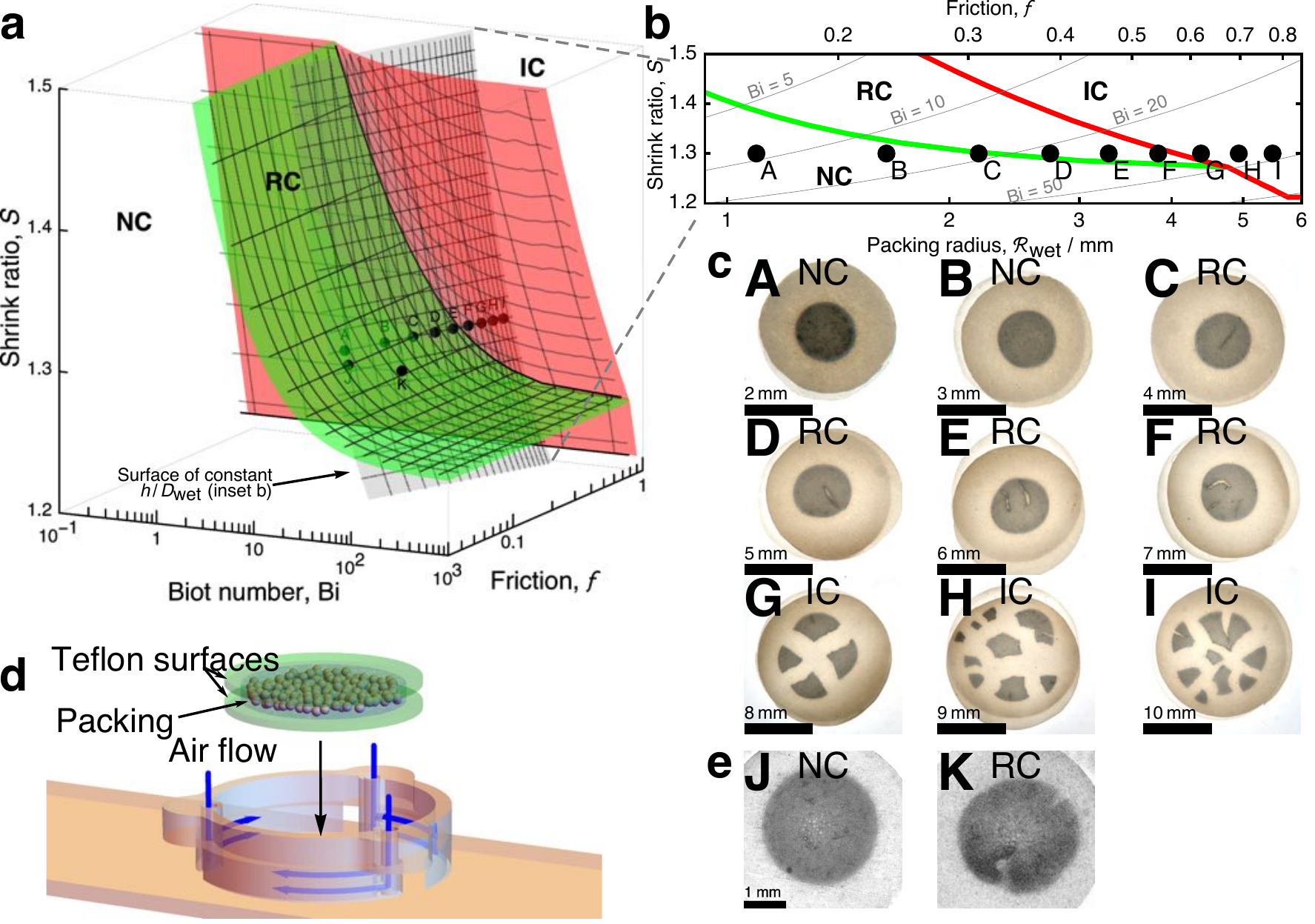}
\caption{Experimentally supported universal description of cracking behavior. (a) A 3D state diagram for the experimentally determined value $\hat{\delta}_\text{wet} = 0.35$ shows the theoretically-determined cracking criteria (colored surfaces) and regions of different cracking behaviors (NC, RC, IC). The green surface in (a) and line in (b) is the NC--RC criterion (Eq.~\ref{eqn:BiRC},\ref{eqn:Smin}); the red surface in (a) and line in (b) represents the NC--IC and RC--IC criteria (Eqs.~\ref{eqn:SIC},\ref{eqn:BiIC},\ref{eqn:Smin}). The gray surface in (a) represents a surface of constant $h/D_\text{wet}$ and is the domain of (b). Packings of different packing radius, $\mathcal{R}_\text{wet}$, represent different points on this gray surface. (c) Drying experiments with packings of varying radius (A--I) show precise agreement with the theoretically-determined cracking criteria. To test the NC--RC crack criterion by varying mass transfer coefficient, (d) a test cell was designed and 3D printed where air flow could be modulated to change the mass transfer coefficient. (e) Two experiments (J,K) below and above the NC--RC border are in agreement with the theory.}
\label{fig:fig5}
\end{figure*}

\subsection{Determining cracking criteria from continuum theory}\label{sec:criteria}
Having verified that the four state variables---$S$, $\text{Bi}$, $\hat{\delta}_\text{wet}$, and $f$---predict cracking, we then used our continuum theory to determine expressions for the borders, or cracking criteria, that delineate NC, RC, and IC behaviors. The criteria arise from different balances of three different stresses: the drying-induced stress from differential shrinkage, $\tilde{\sigma}_\text{max}$ (Eq.~\ref{eqn:genericstress}), the cohesive stress from capillary bridges, $\sim\hat{F}_\text{cap}/\hat{R}_\text{wet}^2$, and the friction-induced stress to immobilize clusters of beads, $\sim f / \hat{R}_\text{wet}^2$. The resulting criteria depend solely on the state variables, demonstrating that these variables fully describe cracking of shrinkable, drying packings. The criteria are plotted in Fig.~\ref{fig:fig4} and show excellent agreement with the results of our DEM simulations.


When friction is small, the NC--RC border is delineated by a threshold $\text{Bi}_\text{RC}$ (blue--green boundaries in Fig.~\ref{fig:fig4}). RC occurs when the $\text{Bi}$-dependent shrinkage stress overcomes the capillary stress, $\tilde{\sigma}_\text{max} \sim \hat{F}_\text{cap,dry}/\hat{R}_\text{wet}^2$; using Griffith's criterion yields similar results (Supplementary Discussion). Balancing these stresses, expressing both sides purely in terms of state variables---$\tilde{\sigma}_\text{max}$ in terms of $S$, $\hat{\delta}_\text{wet}$, and $\text{Bi}$ (Eq.~\ref{eqn:genericstress}); $\hat{F}_\text{cap,dry}$ in terms of $S$ and $\hat{\delta}_\text{wet}$ (Eq.~\ref{eqn:cap})---and solving for $\text{Bi}$ yields an expression for the $\text{Bi}_\text{RC}$ cracking criterion in terms of the state variables:
\begin{equation}
  \text{Bi}_\text{RC} = \left(-0.1-2.22\hat{\mathcal{E}} \epsilon_\text{dry} S^{31/8}/\hat{\delta}_\text{wet}^{1/2}\right)^{-1}
  \label{eqn:BiRC}
\end{equation}
where $\hat{\mathcal{E}} \equiv \mathcal{E}/\left(2\pi\gamma R_\text{wet}\right)$ and the numeric constants were determined from DEM simulation results. $\text{Bi}_\text{RC}$ decreases with increasing $S$ and decreasing $\hat{\delta}_\text{wet}$. This NC--RC criterion demonstrates that a packing is more likely to crack when it is rapidly dried ($\text{Bi}>\text{Bi}_\text{RC}$) and composed of more shrinkable, stiffer beads.

When friction is appreciable, the NC--IC border is delineated by a threshold $S_\text{IC}$ (blue--red boundaries in Fig.~\ref{fig:fig4}). IC occurs when the characteristic friction-induced stress to immobilize clusters in the low $\text{Bi}$ limit overcomes the capillary stress, $f/\hat{R}_\text{wet}^2 \sim \hat{F}_\text{cap,dry}/\hat{R}_\text{wet}^2$; using Griffith's criterion yields similar results \cite{Flores:2017js,2018arXiv180706126M} (Supplementary Discussion). Balancing these stresses and expressing both sides purely in terms of state variables---$\hat{F}_\text{cap,dry}$ in terms of $S$ and $\hat{\delta}_\text{wet}$ (Eq.~\ref{eqn:cap})---and solving for $S$ yields the shrink ratio above which IC must occur due to cluster breakup:
\begin{equation}
  S_\text{IC} = 1.35 \frac{\hat{\delta}_\text{wet}^{4/31}}{f^{8/31}}
  \label{eqn:SIC}
\end{equation}
where the numeric constant was determined from DEM simulation results. $S_\text{IC}$ decreases with decreasing $\hat{\delta}_\text{wet}$ and increasing $f$. This criterion demonstrates that a packing is more likely to break up into clusters when it has more friction with the substrate.

 At high $\text{Bi}$ and high friction, the RC--IC border is delineated by a threshold $\text{Bi}_\text{IC}$ (green--red boundaries in Fig.~\ref{fig:fig4}). IC occurs when the $\text{Bi}$-dependent stress, $\hat{\sigma}_\text{max}$, overcomes the friction-induced stress to immobilize a cluster of beads from the rest of the packing, $\propto f / \hat{R}_\text{wet}^2$. This friction-induced stress also depends on $S$ for large $\text{Bi}$; we used our DEM simulations to empirically determine this dependence with the constraint that $S_\text{IC}$, $\text{Bi}_\text{RC}$, and $\text{Bi}_\text{IC}$ intersect at a point when $S = S_\text{IC}$. Balancing these stresses, expressing both sides purely in terms of state variables---$\tilde{\sigma}_\text{max}$ in terms of $S$, $\hat{\delta}_\text{wet}$, and $\text{Bi}$ (Eq.~\ref{eqn:genericstress})---and solving for $\text{Bi}$ yields the RC--IC criterion in terms of the state variables:
\begin{equation}
  \text{Bi}_\text{IC} = \left(-0.1 - 0.192\hat{\mathcal{E}}f^{65/31}S^{12}\epsilon_\text{dry}/\hat{\delta}_\text{wet}^{48/31}\right)^{-1}
  \label{eqn:BiIC}
\end{equation}
where the numeric constants were determined from DEM simulation results. $\text{Bi}_\text{IC}$ decreases with increasing $S$, decreasing $\hat{\delta}_\text{wet}$, and increasing $f$. This criterion demonstrates that a packing is more likely to break up into clusters when it is rapidly dried ($\text{Bi}>\text{Bi}_\text{IC}$), composed of more shrinkable, stiffer beads, and has more friction with the substrate.

However, with very soft beads, cracking can be completely suppressed even if $S > S_\text{IC}$ (Fig.~S7). There is a $S_\text{NC}$ below which NC must occur due to beads being unable to separate from high precompression. When $S < S_\text{NC}$, the maximum distance that bead radii shrink by, $R_\text{wet} - R_\text{dry}$, does not exceed the distance they are initially precompressed by, $\delta_\text{wet}/2$. By equating these distances, we obtain an expression for $S_\text{NC}$:
\begin{equation}
  S_\text{NC} = \frac{1}{1 - \hat{\delta}_\text{wet}/2} \text{,}
  \label{eqn:Smin}
\end{equation}
which decreases with decreasing $\hat{\delta}_\text{wet}$. This criterion demonstrates that a packing is more likely to crack when it is composed of more shrinkable, stiffer beads.

\begin{figure}[h!]
\centering
\includegraphics[width=\linewidth]{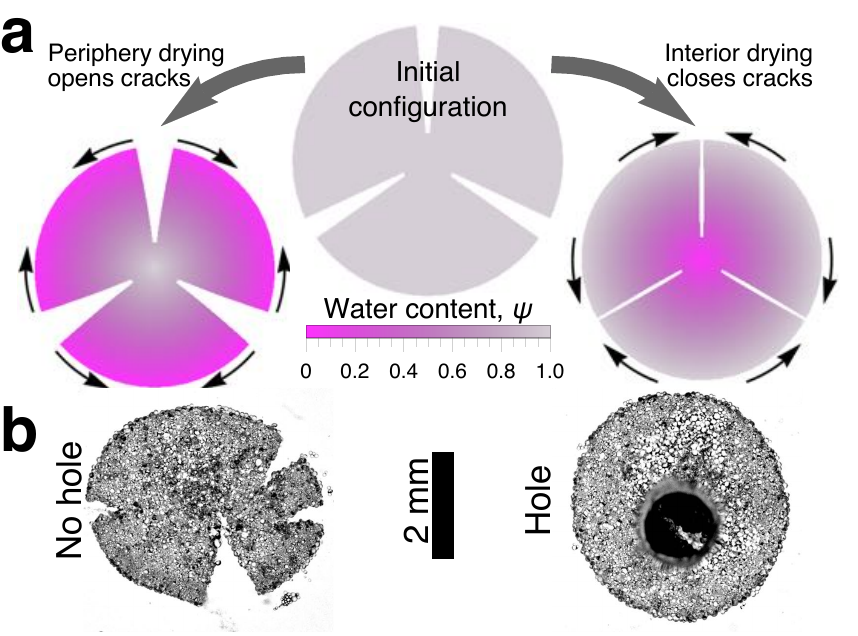}
\caption{Drying profile guides crack evolution. (a) Finite-element solution of our continuum model, applied to a pre-cracked geometry (center). Periphery drying promotes crack opening (left) while interior drying promotes crack self-closing (right). (b) We altered the drying profile experimentally by incorporating a hole in the top surface overlying the packing interior. Periphery drying without the hole leads to cracking at the periphery (left), while drying with the hole, which promotes interior drying, prevents cracking at the periphery (right).}
\label{fig:fig6}
\end{figure}

The combination of the cracking criteria---$\text{Bi}_\text{RC}$, $\text{Bi}_\text{IC}$, $S_\text{IC}$, and $S_\text{NC}$---provides a universal state space of cracking behavior for shrinkable, granular packings. For any packing where the four state variables are known, we can predict how the packing will crack as shown for several values of $\hat{\delta}_\text{wet}$ and $f$ in Fig.~\ref{fig:fig4}. To more completely visualize this universal state space, we plotted the cracking criteria that delineate NC, RC, and IC behaviors for the experimentally determined value of $\hat{\delta}_\text{wet}$ (Fig.~\ref{fig:fig5}a). As an experimental test of the validity of these criteria, we performed multiple drying experiments traversing all three different cracking behaviors in this state space (points A--I in Fig.~{\ref{fig:fig5}}). We first systematically increased packing radius, which simultaneously increases $\text{Bi}$ and $f$. We observed a transition from NC to RC to IC in good agreement with our criteria (Fig.~\ref{fig:fig5}b,c). To further isolate the role of $\text{Bi}$, we performed additional experiments where we varied the mass transfer coefficient, $h$, which independently changes $\text{Bi}$ for a fixed packing size; this was done using a custom 3D-printed test apparatus with precise air-flow control (Fig.~\ref{fig:fig5}d). Changing only $\text{Bi}$, we observed an NC--RC transition in good agreement with the criteria (points J,K in Fig.~\ref{fig:fig5}). These experimental agreements provide support for our description of cracking in shrinkable, granular packings.

\subsection{Controlling cracking}\label{sec:control}
Our experiments (Figs.~\ref{fig:fig1},\ref{fig:openclose}) and experimentally-supported cracking theory demonstrate the central role of differential shrinkage in governing cracking behavior. For the case of drying from the periphery considered heretofore, increasing the drying rate relative to intra-packing transport---as quantified by the Biot number---amplifies differential shrinkage (Fig.~\ref{fig:fig1}). Thus, varying the drying rate is a way to control cracking behavior for periphery drying.

However, our theory is general and not limited to the case of periphery drying. In fact, we can impose any arbitrary shrinkage profile into our continuum model with a pre-cracked geometry and solve for the deformation. To illustrate the influence of different shrinkage profiles, we tested two contrasting cases. When imposing faster drying at the periphery, cracks open as we expect (Fig.~\ref{fig:fig6}a, left); conversely, for the opposite case of faster drying in the interior, cracks close (Fig.~\ref{fig:fig6}a, right). This observation explains the different stages of reversible cracking we found experimentally: cracking, followed by self-closing (Fig.~\ref{fig:fig1}g). At the early stage, periphery beads shrink faster, causing cracks to open; at the later stage, the interior beads shrink faster, decreasing the differential shrinkage and causing cracks to self-close (Fig.~\ref{fig:openclose}). Moreover, this observation suggests a simple method to prevent cracking at the periphery: by promoting interior drying at all times. A direct experimental demonstration of this crack prevention strategy is shown in Fig.~\ref{fig:fig6}b. To promote interior drying, we incorporated a hole into the top surface overlying the interior of a packing. Without the hole, cracks form at the periphery (Fig.~\ref{fig:fig6}b, left); by contrast, with the hole, cracks are suppressed (Fig.~\ref{fig:fig6}b, right). These results show that, in addition to varying the periphery drying rate, another way to control cracks is to vary the spatial profile of drying itself.

\section{Conclusions}
Our work describes the first investigations of how grain shrinkage impacts the cracking of granular packings during drying. Experiments reveal the diverse forms of cracking---shrinkage without cracking, irreversible cracking, and most remarkably, reversible cracking---that can arise due to differential shrinkage. DEM simulations capture these experimental observations, indicating that different macroscopic cracking behaviors can be understood by considering the interplay between bead shrinkage, inter-bead water transport, and inter-bead forces during drying. Moreover, we developed a continuum model that can describe the ultimate state of cracking. As a first step toward a more complete theoretical framework, we focused on a version of the continuum model that uses a linear description of elasticity and does not consider size-dependent changes in bead poromechanical properties; incorporating further details of granular water transport and forces will be a useful direction for future work. Nevertheless, our continuum description sheds light on the underlying physics through the four state variables $S$, $\text{Bi}$, $\hat{\delta}_\text{wet}$, and $f$, which quantify how grain-scale processes determine cracking.

This knowledge can inform applications that need to avoid cracking. It could also suggest ways to control cracking; emerging examples include micro/nano-patterning \cite{Kim:2016bh} and transport in fuel cells \cite{Kim:2016gr}. Moreover, our observation of shrinkage-driven crack evolution suggests a new way to engineer materials that can actuate \cite{Reyssatrsif20090184} and can even self-close, potentially enabling self-healing. Importantly, this approach to self-healing would not require specialized material architectures or additional cross-linking chemicals as current approaches do.

The work presented here focused on differential shrinkage due to drying. However, differential shrinkage can be controlled through other means: for example, by removing or injecting water at targeted points throughout the packing, by imposing osmotic stresses \cite{Datta7041} or changes in solvent quality, and by designing packings with intrinsic, spatially-varying grain transport characteristics. Our results can therefore be generalized beyond the case of drying.

\section{Materials and Methods}
\label{sec:matmet}
\subsection{Experimental}
\subsubsection{Drying experiments}
Drying experiments were performed at $\approx \SI{21}{\celsius}$ with Sephadex G-50F ($R_\text{wet} = \left(67 \pm 12\right) \si{\micro\meter}$). All comparative experiments were performed at the same time to ensure the same humidity and drying rate. Hydrophobic surfaces in the results presented in Fig.~\ref{fig:fig1} and Supplementary Videos~S1--S3 were circular siliconized glass coverslips, which were either purchased (Hampton) or silanized (see below). Hydrophobic surfaces in the results presented in Figs.~\ref{fig:fig5},\ref{fig:fig6} were made from \SI{51}{\micro\meter}-thick PTFE. Circles were cut out using a razor cutter (Roland STIKA SV-8).

To create disk-shaped packings, hydrogels were hydrated in centrifuge tubes with deionized water. After hydrogels settled to the bottom of the tube after approximately five minutes, excess water was removed such that the water level was a few millimeters above the packing. To ensure that the deposited suspension had a packing density close to random close packing, pipette tips were submerged several millimeters below the top of the packing during bead when filling. Droplets were then deposited from the pipette to circular hydrophobic surfaces. Another hydrophobic surface (top surface) was then placed on top of the droplet; it therefore follows the top of the packing throughout the drying process. The droplet suspension then expanded such that its contact lines became pinned to the circumferential edges of the top and bottom surfaces, rendering a disk-like geometry.

Fluorescence micrographs (Figs.~\ref{fig:fig1},\ref{fig:openclose},\ref{fig:modelvalfig} were obtained via confocal microscope (Nikon A1R) with the widest pinhole settings. Fluorescein sodium salt in deionized water (\SI{0.1}{\micro\gram\per\milli\liter}) was used to fluorescently label hydrogels. Samples were excited at \SI{488}{\nano\meter} and imaged at \SIrange{500}{550}{\nano\meter}. Conventional photographs (Figs.~\ref{fig:fig5},\ref{fig:fig6}) were obtained via CMOS camera (Sony A6300) with a macro lens (Nikon Micro-Nikkor 105mm f/2.8 AF-D).

Experiments where the mass transfer coefficient (drying speed) was varied were performed on a custom 3D-printed device (Fig.~\ref{fig:fig5}d). A packing between two hydrophobic surfaces was placed in the device. An annular region provided airflow in a clockwise direction, which was supplied by air channels. These air channels were connected to lab air supply and controlled via a flow regulator to rates ranging \SIrange{0}{600}{\liter\per\hour}.

\subsubsection{Glass silanization procedure}
\SI{1.97}{\micro\liter\per\milli\liter} of trichloro(octadecyl)silane was added to a mixture of 80\% dodecane and 20\% chloroform by volume. \SI{5}{\milli\meter} round coverslips were made hydrophobic by submersing them in the solution and mixing for 20 minutes. The coverslips were then rinsed with chloroform, acetone, and finally with water.

\subsubsection{Hydrogel bead property measurements ($R_\text{wet}$, $S$, $D_\text{wet}$)}
Wet bead radius, $R_\text{wet}$, shrink ratio, $S$, and poroelastic diffusion coefficient, $D_\text{wet}$ of Sephadex G-50F were measured by imaging the time-dependent swelling process of individual beads. Individual beads were placed on a glass slide under observation of a confocal microscope. Then, a droplet of deionized water was placed in contact with the beads to swell the beads. The time-series images of swelling were processed using Mathematica\textregistered. An edge detection filter was applied to obtain circles around each bead. Then, a circle Hough transform was applied to obtain the locations and sizes of circles. Identical beads were identified across successive frames by finding similarly positioned and sized beads. The wet bead size was measured to be $R_\text{wet} = \left(67 \pm 12\right) \si{\micro\meter}$ (uncertainties are one standard deviation). By comparing bead sizes before and after swelling, $S = 1.30 \pm 0.16$. The poroelastic diffusion coefficient was determined by fitting the 1D spherical diffusion equation solution with Dirichlet boundary condition to the time-dependent bead volume data. This diffusion coefficient was set to $D_\text{wet}$. This coefficient was measured to be $D_\text{wet} = \left(8.6 \pm 6.2\right) \times 10^{-7} \si{\centi\meter\squared\per\second}$. Uncertainties in $R_\text{wet}$, $S$, and $D_\text{wet}$ are likely due to imprecisions in image processing and variations in the manufacturing process. For modeling purposes, $R_\text{wet} = \SI{67}{\micro\meter}$, $S = 1.30$, and $D_\text{wet} = \SI{8.6e-7}{\centi\meter\squared\per\second}$.

\subsubsection{Friction measurement ($\hat{F}_\text{fr}$)}
Friction for PTFE surfaces was measured by first placing a packing with a volume of several hundred \si{\micro\liter} on a PTFE surface taped to a glass slide. The weight of the packing was measured on a scale. Then, the surface was tilted and the angle at the moment when the packing began to slide was recorded (the moment when static friction is overcome). The contact area with the surface was determined by photographing the underside of the slide (the PTFE was slightly transparent). Using the weight of the packing, and the angle of initial sliding, the static friction force was determined. Dividing this static friction force by the average area per bead (assuming hexagonal close packing and $R_\text{wet} = \SI{67}{\micro\meter}$) yields the static friction threshold for a single bead: $\hat{F}_\text{fr} = \left(0.054 \pm 0.019\right)$. For modeling purposes, we used $\hat{F}_\text{fr} = 0.053$.

\subsubsection{Precompression measurement ($\hat{\delta}_\text{wet}$)}
Precompression ($\hat{\delta}_\text{wet}$) was measured using a fluorescent micrograph of a packing at the wet state. By measuring the length of lines between beads ($\approx 2a$), $\delta$ can be determined using Eq.~\ref{eqn:capbridgeradius}. A value of $\hat{\delta}_\text{wet} = \left(0.36 \pm 0.11\right)$ was measured (uncertainty is one standard deviation). For modeling purposes, $\hat{\delta}_\text{wet} = 0.35$.

\subsubsection{Biot number measurement ($\text{Bi}$)}
The Biot number was determined by measuring the velocity of packing shrinkage, $v_\text{wet} = \text{Bi}\left(1 - \hat{V}_\text{dry}\right)\left(-\epsilon_\text{dry}\right)D_\text{wet}/\mathcal{R}_\text{wet}$. Time-series images of a packing shrinking over time during the air invasion process showed nearly constant disk shrinkage velocity ($\mathcal{R}$ linearly decreasing with time) up to the conclusion of air invasion. The moment when air invasion ends and individual beads begin to shrink is a state we define as the wet state. However, in some cases beads started to shrink before air invasion was completed. Thus, the wet state is an idealization for modeling purposes. It is a reasonable idealization since any overlap time between shrinkage and air invasion was short compared to the length of the experiment.

\subsection{Scaling relationships for poromechanical properties}\label{sec:scaling}
As a hydrogel bead shrinks, its hydrated polymer mesh deswells. We treat this change by assuming that the mesh size $\xi \propto V^{3/4}$, as experimentally verified by others for many hydrogels, including those used in our experiments, for which the poromechanical properties can be described by treating the hydrogel as a semi-dilute polymer solution \mbox{\cite{deGennes:1979uw,vanderSman:2015ef,ZRINYI19871139,doi:10.1021/ma011408z}}. Using scaling relationships developed by de Gennes \cite{deGennes:1979uw}, the bulk modulus, $K$, is related to the mesh size, $\xi$, at constant temperature by $K \propto \frac{1}{\xi^3}$, and thus, $K \propto V^{-9/4}$. In terms of the wet state, we can write an explicit power-law dependence of the modulus on the state of swelling ($\hat{V}$ or $\hat{R}$):
\begin{equation}
	K = K_\text{wet} \hat{V}^{-9/4} = K_\text{wet} \hat{R}^{-27/4}\text{.}
\end{equation}
Since at any intermediate state $\hat{R} \sim S^{-1}$, we can describe the modulus in terms of $S$. Nondimensionally,
\begin{equation}
	\hat{K} \equiv \frac{K}{2\pi\gamma / R_\text{wet}} = \hat{K}_\text{wet} \hat{V}^{-9/4} = \hat{K}_\text{wet} \hat{R}^{-27/4} \sim \hat{K}_\text{wet} S^{27/4}
\label{eqn:Khat}
\end{equation}
where we nondimensionalize forces by $2\pi\gamma R_\text{wet}$ and lengths by $R_\text{wet}$; thus, we nondimensionalize pressure by $2\pi\gamma R_\text{wet} / R_\text{wet}^2 = 2\pi\gamma / R_\text{wet}$.

Similarly, the single-bead permeability $\kappa$ is a function of the state of swelling for an individual bead: $\kappa \propto \xi^2$ and thus $\kappa \propto V^{3/2}$. We can again write an explicit power-law dependence of the permeability on shrink ratio in terms of the wet state:
\begin{equation}
	\kappa = \kappa_\text{wet} \hat{V}^{3/2} = \kappa_\text{wet} \hat{R}^{9/2} \sim \kappa_\text{wet} S^{-9/2} \text{.}
	\label{eqn:kappa}
\end{equation}

Since the poroelastic diffusion coefficient, $D$, depends on both $K$ and $\kappa$ where $D \equiv \frac{\kappa K}{\mu \hat{V}}$ (Section~\ref{sec:diffusion}), the diffusion coefficient also scales with the shrink ratio:
\begin{equation}
	D \equiv \frac{\kappa K}{\mu \hat{V}} = \frac{\kappa_\text{wet} K_\text{wet}}{\mu} \hat{V}^{-7/4} = \frac{\kappa_\text{wet} K_\text{wet}}{\mu} \hat{R}^{-21/4} \sim \frac{\kappa_\text{wet} K_\text{wet}}{\mu} S^{21/4}
\end{equation}

The compression distance, $\delta$, and its nondimensional form $\hat{\delta}_\text{wet} \equiv \delta/R_\text{wet}$, are also a function of the state of swelling since $\hat{K}$ is a function of the state of swelling:
\begin{equation}
  \hat{\delta} = \frac{4}{3\hat{K}} = \frac{4\hat{V}^{9/4}}{3\hat{K}_\text{wet}} =\frac{4\hat{R}^{27/4}}{3\hat{K}_\text{wet}} \sim \frac{4S^{-27/4}}{3\hat{K}_\text{wet}} \text{.}
  \label{eqn:delta}
\end{equation}

\subsection{DEM simulations}
\subsubsection{Algorithm}\label{sec:algorithm}
Networks of individual shrinkable beads were simulated using the equations of discrete transport and discrete force interactions. The packing geometry was a circular 2D monolayer of hexagonally-close-packed spherical beads with bonds between adjacent beads. Simulations were performed using Mathematica. The simulation procedure was as follows.
\begin{enumerate}
  \item Calculate bead sizes for a new timestep using discrete bead-to-bead transport relations (Eqs.~\ref{eqn:discretetransportvol},\ref{eqn:demdiffcoeff},\ref{eqn:discreteboundaryflow}--\ref{eqn:volchange}) ensuring that the timestep is small enough such that bead radii do not change more than \SI{0.1}{\percent}. To find an optimal timestep size, we used a root-finding algorithm.
  \item Perform a gradient descent where beads are moved in timeless steps (quasistatic relaxation) in the direction of their net forces. These forces are calculated using discrete contact forces (Eq.~\ref{eqn:contactforce}) and capillary forces (Eq.~\ref{eqn:cap}). Continue gradient descent until the smallest force on a bead in the system is less than 1\% of the force to break a bond.
  \item If, after gradient descent, any bonds are overstretched, check whether the amount of overstretching exceeds 1\% of $\delta_\text{wet}$ (hybrid event detection)---using smaller values of excessive overstretching did not greatly affect DEM simulation results. If it does, then go back to Step 1 with half the timestep size and proceed in time (Steps~1--3) until the same cracking event is reached. If the amount of overstretching still exceeds 1\% of $\delta_\text{wet}$, go back to the last timestep where cracking did not occur with half the timestep size. Repeat with finer and finer timestep sizes until the amount of overstretching is less than 1\% of $\delta_\text{wet}$. Once the overstretching is small enough, break the bond, and perform a gradient descent.
   \item Continue timestepping (Steps~1--3) until all beads have reached their final dry size.
\end{enumerate}

\subsubsection{Water transport and evaporation}\label{sec:demflow}
We discretize Darcy's law
between two beads $i$ and $j$ such that the volumetric flow rate between them is given by $Q_{ij} = \pi a_{ij}^2 \frac{\kappa_{ij}}{\mu} \frac{P_i - P_j}{R_i+R_j-\delta_{ij}}$, where $a_{ij}$ is the radius of the capillary bridge linking the beads, $\kappa_{ij}$ is an effective permeability for the pair of beads, $\mu$ is the dynamic viscosity of water, $P_{i}$ and $R_{i}$ are the liquid pore pressure and radius of bead $i$, and $\delta_{ij}$ is the precompression distance between the beads determined from Hertzian mechanics (Fig.~{\ref{fig:fig1}}e). 

We use poroelasticity to convert the gradient in $P$ to a gradient in bead volumes {\cite{goehring2015desiccation,PhysRevApplied.6.064010}}. Specifically, the bulk modulus $K=-VdP/dV$, and thus,  $P_{i}-P_{j}=\frac{K_{ij}}{\hat{V}_{ij}V_{\text{wet}}}(V_{i}-V_{j})$, where $\hat{V}_{ij}$ is an effective nondimensional bead volume. Thus,
\begin{equation}
  Q_{ij} = \pi a_{ij}^2 \frac{D_{ij}}{V_\text{wet}}\frac{V_i - V_j}{R_i+R_j-\delta_{ij}} \text{.}
  \label{eqn:discretetransportvol}
\end{equation}
The discretized diffusion coefficient is 
\begin{equation}
  D_{ij} = \text{min}\left(D_i,D_j\right)
  \label{eqn:demdiffcoeff}
\end{equation}
where
\begin{equation}
	D_i \equiv \frac{\kappa_i K_i}{\mu \hat{V}_i}
\end{equation}
is a mutual poroelastic diffusion coefficient---assuming transport between the beads is limited by the slower diffusing of the two (detailed further in Section~\ref{sec:diffusion}). Our DEM simulations use this discretized form of Darcy's law with size-dependent $D_{ij}$, as described by the scaling relationships in {Section~\ref{sec:scaling}. 

The discretized flow at the periphery of the packing is
\begin{equation}
  Q_{\text{evap},i} = A_{\text{evap},i} \left(\text{Bi} \frac{D_\text{wet}}{\mathcal{R}_\text{wet}}\right) \left(-\epsilon_\text{dry}\right)\left(\hat{V} - \hat{V}_\text{dry}\right) \text{.}
  \label{eqn:discreteboundaryflow}
\end{equation}
where $A_{\text{evap},i}$ the evaporative area for a single periphery bead $i$, $\text{Bi}$ is the Biot number,  $D_\text{wet} = \kappa_\text{wet} K_\text{wet}/\mu $ is the poroelastic diffusion coefficient evaluated at the wet state, $\mathcal{R}_\text{wet}$ is the initial packing radius, and $\epsilon_\text{dry}$ is given in Eq.~\ref{eqn:edry}. For the purpose of approximating $A_{\text{evap},i}$, we assume periphery beads are cylindrical with a circumference of $2\pi\left(R_i- \frac{\delta_{ii}}{2} \right)$ and a height of $2\left(R_i- \frac{\delta_{ii}}{2} \right)$. The evaporative area for a single periphery bead is then
\begin{equation}
  A_{\text{evap},i} = 4 \omega_i \pi \left(R_i- \frac{\delta_{ii}}{2} \right)^2
\end{equation}
where $\omega_i$ is the 2D angular fraction of the bead that is visible to the environment (line of sight). This angular fraction is calculated via $\omega_i = 2\pi - \alpha_i / \left(2\pi\right)$ where $\alpha_i$ is the total field of view from every other bead. The total volume change for bead $i$ over a timestep size of $\Delta t$ is
\begin{equation}
  \Delta V_i = -\left(\sum_{j \in N_i} Q_{ij} + Q_{\text{evap},i}\right) \Delta t \label{eqn:volchange}
\end{equation}
where $Q_{ij}$ is determined from Eq.~\ref{eqn:discretetransportvol}.

\subsubsection{Force interactions}\label{sec:demforces}
Motivated by previous work {\cite{Moller:2007bp,Fogden:1990ej,Butt:2010gs,Herminghaus:2005ei}}, and validated for hydrogel beads {\cite{PhysRevE.84.011302,PhysRevE.82.041403}}, we model the contact force using a simple Hertzian mechanical description:
\begin{equation}
  F_{\text{con},ij} = \frac{4}{3}\bar{E}_{ij} \bar{R}_{ij}^{1/2} d_{ij}^{3/2}
  \label{eqn:contactforce}
\end{equation}
where $d_{ij}$ is the instantaneous deformation distance and the bead moduli and sizes are given by
\begin{align}
  \frac{1}{\bar{E}_{ij}} &= \left(\frac{1-\nu^2}{E_i}+\frac{1-\nu^2}{E_j}\right) \label{eqn:eij}\\
  \frac{1}{\bar{R}_{ij}} &= \frac{1}{R_i}+\frac{1}{R_j}.
\end{align} We assume that Poisson's ratio, $\nu$, is equal to $1/3$ under quasistatic conditions and thus $E = K$, as measured previously for cross-linked hydrogels \cite{C003344H,doi:10.1021/ma50002a038,Andrei1998}. Motivated by our observations of thin annular capillary bridges (Fig.~{\ref{fig:fig1}}c,d) and other modeling work \cite{Moller:2007bp,doi:10.1021/la000657y}, we consider an attractive capillary force between the beads:
\begin{equation}
  F_{\text{cap},ij} = 2\pi a_{ij} \gamma
  \label{eqn:cap}
\end{equation}
where $\gamma$ is the surface tension of water and
\begin{equation}
  a_{ij} = \sqrt{\delta_{ij}\bar{R}_{ij}}
  \label{eqn:capbridgeradius}
\end{equation}
is the radius of contact between beads. Because the hydrogel beads are hydrated polymer networks---containing a continuous fluid phase through each bead---and thus, the size of the hydrated polymer network in each bead is contiguous with the fluid volume, we assume this contact radius is the capillary bridge annular radius. Once this capillary force is overcome, the capillary bridge link between two spheres is broken. At mechanical equilibrium when $F_{\text{con},ij} = F_{\text{cap},ij}$, $d_{ij} = \delta_{ij}$ and
\begin{equation}
  \delta_{ij} = \frac{3\pi\gamma}{2\bar{K}_{ij}} \text{.}
  \label{eqn:dimensionaldelta}
\end{equation}
and thus $\hat{\delta}_{ij} \equiv \delta_{ij}/R_\text{wet} = \left(\hat{\delta}_\text{wet} /2\right) \left( \hat{K}_\text{wet}/\hat{K}_i + \hat{K}_\text{wet}/\hat{K}_j \right)$.

\subsubsection{Quasistatic assumption}\label{sec:demquasistatic}
The quasistatic assumption is valid since the diffusion timescale is much smaller than the mechanical relaxation timescale. Using a very approximate mobility of \SI{e5}{\meter\per\second\per\newton} (extracted from Fig.~5 of \cite{doi:10.1021/la2018083}), a characteristic of force $2\pi\gamma R_\text{wet}$ ($\gamma = \SI{72}{\milli\newton\per\meter}$, $R_\text{wet}=\SI{67}{\micro\meter}$), and a characteristic system length scale of \SI{1}{\milli\meter}, a characteristic mechanical relaxation timescale is $\sim \SI{1}{\milli\second}$. Conversely, using $D_\text{wet} = \SI{8.6e-7}{\centi\meter\squared\per\second}$ and the same system length scale of \SI{1}{\milli\meter} gives a characteristic time that is $\sim \SI{e4}{\second}$. As a result, the diffusion is many orders of magnitude slower than mechanical relaxation.

\subsubsection{Cracking behavior classification}\label{sec:demcracking}
DEM results were classified as either NC, RC, and IC according to a set of rules. NC occurs if the number of bond breaks is less than 1\% of the total number of bonds. IC occurs if, in the final state, the standard deviation of bond distances is more than 10\% of the mean bond distance, or if the average number of bond breaks that occur after the point of maximum strain energy is more than three times the number of bond breaks at the point of maximum strain energy. Otherwise, RC occurs.

\subsection{Derivation of the diffusive transport model} \label{sec:diffusion}
We can write a statement of conservation of mass (incompressible fluid) for a single bead with $N$ bridges attached (a coordination number of $N$).
\begin{equation}
	\frac{\partial V}{\partial t} = \sum_i^N \iota_{\text{in},i} A_\text{bridge}
\end{equation}
where $\iota_{\text{in},i}$ is the inward volumetric flux of water through the $i$th bridge and $A_\text{bridge}$ is the cross-sectional area of a liquid bridge. Around each bead, let us consider a volume called the \emph{Voronoi cell}, $\mathcal{V}$, which is the space-filling volume associated with each bead. Thus, the Voronoi cells can be tessellated with each other. The bounding surface of each Voronoi cell is what we will call $S_\text{cell}$. The with outward facing normal associated with $S_\text{cell}$ is $n$. With these definitions, we can express the right hand side as a closed surface integral around $S_\text{cell}$.
\begin{equation}
	\sum_i^N \iota_{\text{in},i} A_\text{bridge} = -\oiint_{S_\text{cell}} \left(\vectorsym{\iota} \cdot \vectorsym{n}\right) dS
\end{equation}
The divergence theorem states that
\begin{equation}
	\oiint_{S_\text{cell}} \left(\vectorsym{\iota} \cdot \vectorsym{n}\right) dS = \iiint_{\mathcal{V}}\left(\nabla \cdot \vectorsym{\iota}\right) dV \text{;}
\end{equation}
thus,
\begin{equation}
	\frac{\partial V}{\partial t} = -\iiint_{\mathcal{V}}\left(\nabla \cdot \vectorsym{\iota}\right) dV \text{.}
\end{equation}
Since the average of the flux divergence within the bead is related to the volume integral by
\begin{equation}
	\left(\nabla \cdot \vectorsym{\iota}\right)_\text{avg} = \frac{1}{\mathcal{V}}\iiint_{\mathcal{V}}\left(\nabla \cdot \vectorsym{\iota}\right) dV \text{,}
\end{equation}
we can express the change in mass as
\begin{equation}
	\frac{\partial V}{\partial t} = -\left(\nabla \cdot \iota\right)_\text{avg} \mathcal{V} \text{.}
\end{equation}
where $t$ is time. Assuming a bead can be treated as infinitesimally sized and that there is an effective mean continuum flow field $\vectorsym{j}$ where $\left(\nabla \cdot \vectorsym{\iota}\right)_\text{avg} = \left(\nabla \cdot \vectorsym{j}\right)$, then
\begin{equation}
	\frac{\partial V}{\partial t} = -\left(\nabla \cdot \vectorsym{j}\right)\mathcal{V} \text{.}
\end{equation}
The transport constitutive law for the continuum flow
field $j$ is
\begin{equation}
	\vectorsym{j} = -\frac{\kappa}{\mu}\nabla P = -\frac{\kappa K}{V\mu}\nabla V  \text{.}
	\label{eqn:efftransportconstitutive}
\end{equation}
The quantity $V/\mathcal{V}$ is the local solid fraction that varies in space and time. Reapplying Darcy's law, the poroelastic relation between pressure and volume is given by,
\begin{equation}
	\frac{\partial V}{\partial t}  = \left(\nabla \cdot \left(\frac{\kappa K}{\mu V}\nabla V\right) \right) \mathcal{V} \text{.}
\end{equation}
Multiplying the right hand side by $V_\text{wet}/V_\text{wet}$, we can express the transport equation as
\begin{equation}
	\frac{\partial V}{\partial t} = \left(\nabla \cdot \left(\frac{\kappa K}{\mu}\frac{V_\text{wet}}{V} \nabla V\right) \right) \frac{\mathcal{V}}{V_\text{wet}} = \left(\nabla \cdot \left(D\nabla V\right) \right) \hat{\mathcal{V}}
\end{equation}
where $D \equiv \frac{\kappa K}{\mu}\frac{V_\text{wet}}{V}$. We can nondimensionalize this diffusion equation as
\begin{equation}
	\frac{\partial \hat{V}}{\partial \tau}  = \left(\nabla \cdot \left(\hat{D}\nabla \hat{V}\right) \right) \hat{\mathcal{V}} \label{eqn:fullcontinuumappendix}
\end{equation}
where $\tau \equiv \frac{D_\text{wet}t}{\mathcal{R}_\text{wet}^2}$, $\hat{D} \equiv D/D_\text{wet}$. Eq.~\ref{eqn:fullcontinuumappendix} is a continuum treatment of transport that takes into account size-dependent changes in $D$ (used in Fig~\ref{fig:modelvalfig}). The changes in $D$ can be handled with polymer scaling relationships. Since $\kappa \propto \xi^2$, $K \propto \xi^{-3}$, $\xi \propto V^{3/4}$, and $\hat{V}^{1/3}=\hat{R}\sim \frac{1}{S}$
\begin{equation}
	\hat{D} = \hat{V}^{-7/4} \sim S^{21/4}\text{.}
\end{equation}
The parameter $\mathcal{V}/V_\text{wet}$ is thus the reciprocal of the packing volume fraction at the wet state; as a simplifying assumption, we take this to be $\approx1$. Moreover, because $S=1.3$ is of order unity in the experiments, we focus on a version of the continuum model that does not consider size-dependent changes in bead properties, for simplicity. Indeed, for materials where $S < 1.5$, $\hat{D}$ does not vary more than an order of magnitude; thus, a constant-diffusion transport approximation is reasonable in many cases ($\hat{D} \approx 1$). Thus, the transport equation can be reasonably approximated as
\begin{equation}
	\frac{\partial \psi}{\partial \tau} = \nabla^2 \psi
\end{equation}
for $S < 1.5$. Here, we have used $\psi$, defined as
\begin{equation}
	\psi \equiv \frac{\hat{V}-\hat{V}_\text{dry}}{1-\hat{V}_\text{dry}}\text{,}
\end{equation}
as it spans $0 \leq \psi \leq 1$, which is convenient for transport boundary conditions.

Analytically solving this simplified continuum diffusion equation with the convective boundary condition (Eq.~{\ref{eqn:bc}}), yields Eq.~\ref{eqn:simpletransportsolution}:
\begin{equation*}
  \psi = \sum_{n=1}^\infty C_n e^{-\lambda_n^2 \tau} J_0\left(\lambda_n \tilde{r}\right)
\end{equation*}
where $C_n \equiv \frac{2 J_1\left(\lambda_n\right)}{\lambda_n\left(J_0^2\left(\lambda_n\right)+J_1^2\left(\lambda_n\right)\right)}$, $J_0$ and $J_1$ are the zeroth and first-order Bessel functions of the first kind, $\tau \equiv D_\text{wet} t/\mathcal{R}_\text{wet}^2$ is the Fourier number, and $\lambda_n$ are the roots of $\lambda_n\frac{J_1\left(\lambda_n\right)}{J_0\left(\lambda_n\right)} = \text{Bi}$. This relationship is plotted in Fig.~S2.

\subsection{Derivation of drying-induced stresses} \label{sec:continuumshrinkage}

To determine the stresses due to differential shrinkage, we develop a continuum model using linear elasticity. Specifically, we build a constitutive relationship between stress, $\tensorsym{\sigma}$, and strain, $\tensorsym{\epsilon}$, by adding an extra drying-induced shrinkage term to Hooke's law (Eq.~\ref{eqn:elasticity1}), where the quantity $\epsilon_\text{s}<0$ represents the strain due to bead shrinkage from water evaporation. Specifically, we define $\epsilon_\text{s}$ as the relative change in cross-sectional length per bead due to a change in water content:
\begin{equation}
  \epsilon_\text{s} \equiv \frac{\left(\hat{R}-\hat{\delta}/2\right) - \left(\hat{R}_\text{wet}-\hat{\delta}_\text{wet}/2\right)}{\hat{R}_\text{wet}-\hat{\delta}_\text{wet}/2} =  \frac{\left(\hat{R}-\hat{\delta}_\text{wet}\hat{V}^{9/4}/2\right)}{1-\hat{\delta}_\text{wet}/2} - 1 \text{.}
  \label{eqn:actualshrinkagestrain}
\end{equation}
At the wet state, $\epsilon_\text{s} = 0$ and at the dry state, $\epsilon_\text{s} = \epsilon_\text{dry}$, which is related to the state variables $S$ and $\hat{\delta}_\text{wet}$:
\begin{equation}
  \epsilon_\text{dry} \equiv \frac{\left(\hat{R}_\text{dry}-\hat{\delta}_\text{dry}/2\right) - \left(\hat{R}_\text{wet}-\hat{\delta}_\text{wet}/2\right)}{\hat{R}_\text{wet}-\hat{\delta}_\text{wet}/2} = \frac{1/S - \hat{\delta}_\text{wet}S^{-27/4}/2}{1 - \hat{\delta}_\text{wet}/2} - 1
  \label{eqn:edry}
\end{equation}
where $\hat{R}_\text{dry} = 1/S$ and $\delta_\text{dry} = \delta_\text{wet}S^{-27/4}$. To obtain a simplified and analytical solution for displacement and stress, we linearize Eq.~{\ref{eqn:actualshrinkagestrain}} such that $\epsilon_\text{s}$ is proportional to changes in volumetric water content, $1-\psi$, and retains the same values at the wet and dry states (Eq.~\ref{eqn:shrinkagestrain}):
\begin{equation*}
  \epsilon_\text{s} \approx \epsilon_\text{dry} \left(1 - \psi\right) \text{.}
\end{equation*}
This simplifying assumption does not considerably affect the model results, as exemplified in Fig.~{\ref{fig:modelvalfig}}. Moreover, incorporating this approximation into the constitutive relation (Eq.~{\ref{eqn:elasticity1}}) within the framework of linear elasticity, along with the water transport solution (Eq.~\ref{eqn:simpletransportsolution}), yields the full, analytical time-dependent displacement and stress fields (Eqs.~\ref{eqn:analyticalur},\ref{eqn:stress}):
\begin{equation*}
  \tilde{u}_r = \epsilon_\text{dry}\left(\tilde{r}-\sum_n^\infty \frac{2 C_n}{3 \lambda_n}\left(\tilde{r} J_1\left(\lambda_n\right)+ 2 J_1\left(\lambda_n \tilde{r}\right)\right) e^{-\lambda_n^2 \tau} \right) \text{,}
\end{equation*}
\begin{equation*}
  \frac{\sigma_{\theta\theta}\left(\tilde{r},\tau\right)}{\mathcal{E}} = -\frac{\epsilon_\text{dry}}{\tilde{r}}\sum_n^\infty \frac{C_n}{\lambda_n} \left(\tilde{r} J_1\left(\lambda_n\right)+  J_1\left(\lambda_n \tilde{r} \right) - \tilde{r} \lambda_n J_0\left(\lambda_n \tilde{r}\right) \right) e^{-\lambda_n^2 \tau} \text{.}
\end{equation*}

\section*{Conflicts of interest}
There are no conflicts to declare.

\section*{Acknowledgements}
It is a pleasure to acknowledge S. Hilgenfeldt, L. Mahadevan, A.Z. Panagiotopoulos, W.B. Russel, and G.W. Scherer for stimulating discussions. We also acknowledge the Princeton Institute for Computational Science and Engineering for computer cluster access. This work was supported by start-up funds from Princeton University, the Alfred Rheinstein Faculty Award, the Grand Challenges Initiative of the Princeton Environmental Institute, the ReMatch+ Summer Internship Program, and in part by funding from the Princeton Center for Complex Materials, a Materials Research Science and Engineering Center supported by NSF grant DMR-1420541. 



\balance


\bibliography{refs} 
\bibliographystyle{rsc} 

\end{document}